\documentclass[final,3p,times,twocolumn]{elsarticle}

\usepackage{amsmath}
\usepackage{epstopdf}
\usepackage{aas_macros}

\usepackage[figuresright]{rotating}
\usepackage{subfigure}
\usepackage{graphicx,epsfig}
\usepackage{epstopdf}
\usepackage{float}
\usepackage{mathtools}
\usepackage{epstopdf}
\usepackage{amsmath,amssymb}
\usepackage{color}

\begin{document}

\begin{frontmatter}

\title{Exploring the Voids: Luminosity Functions and Magnetic Field}

\author[First,add]{Agniva Roychowdhury}

\author[Second]{Saumyadip Samui\corref{cor1}}
\ead{saumyadip.physics@presiuniv.ac.in}

\cortext[cor1]{Author for correspondence.}

\address[First]{Department of Physics, University of Maryland Baltimore County, 1000 Hilltop Circle, Baltimore MD 21250.}
\address[add]{Department of Physics,  Presidency University, 86/1 College Street, Kolkata 700073.}
\address[Second]{Department of Physics,  Presidency University, 86/1 College Street, Kolkata 700073.}

\begin{abstract}
We present semi-analytical models for magnetisation of the void inter-galactic medium (IGM) by outflows from void galaxies. The number density of dark matter haloes in an under-dense region (i.e., void) is obtained using the excursion set method extended for such low density environment. The star formation in such haloes has been estimated, taking account of the negative feedback by supernovae. The galaxy formation/evolution
model is tuned to provide the $r$-band luminosity function, the stellar mass function and also the color of void galaxies as obtained from recent observations. This star formation model is used to study possible outflows from void galaxies driven by the hot thermal gas and cosmic ray pressures. These outflows drag the magnetic fields present in those galaxies to the void IGM. We show that
such a model can magnetise $\sim 30\%$ of the void IGM with the magnetic field strength of $10^{-12}-10^{-10}$~G while considering only magnetic flux freezing condition. Along with this, the megapersec size of individual outflows can explain the non-detection of GeV photons in TeV blazars that put a lower limit of $10^{-16}$~G for void IGM magnetic field with a Mpc coherent length scale.
\end{abstract}

\end{frontmatter}
\section{Introduction}

Cosmic voids are the vast regions of nearly empty space spanning even upto few hundreds of megaparsecs, occurring between regions of high density environments like filaments and superclusters. One of the foremost detections of voids in galaxy redshift surveys dates to as early as \cite{kirsh81}, who detected a void of size $\sim$ 70 Mpc. Since then, numerous other galaxy surveys have opened up the window to detect voids in the galaxy distribution
\citep{kauff91,elad96,elad97,hoyle02,platen07}. These structures have been observed in high redshifts as well, and studies that indicate their formation history to be different from their high-density counterparts \citep[i.e.][]{sheth04} provide new areas to probe complimentary galaxy formation theories as a mirror to those already existing.

By definition, voids contain only a few galaxies. Hence, it is difficult to have a correct understanding of them, in particular, how these galaxies are different
from their counterparts in the high density universe.
There has been significant progress in the theoretical domain to study the formation and evolution of voids, wherein both analytic methods and numerical simulations have been invoked to explain the void phenomenon \citep[][]{2006MNRAS.368.1132P,2005ApJ...635..990C}. On the other hand, several observational studies sought to understand the luminosity function, star formation, and subsequently galaxy mass function in cosmic voids  \citep{gold05,beygu16}. While many have discussed statistical properties of voids as a whole  \citep{vonbenda08,hoyle12} and properties of residing galaxies \citep{rojas04,rojas05} and their formation-evolution \citep{croton08}, several works have dedicated themselves to constraining the cosmological parameters using the observations of voids \citep{rijo09}. In spite of these studies concerning voids and its galaxies, many primary questions still remain open, which include how galaxies in voids formed and evolved, and if there is any correlation of void galaxy properties to that of galaxies in the high density universe.

Specifically, in the scenario of galaxy luminosity functions in voids, \cite{croton05} and \cite{croton08}, for example, have
demonstrated that Schechter mass functions can approximately describe various properties of the void galaxy population, both old and young. Their results also included the fact that early and late galaxies evolve very differently for clusters and voids.
Later on, \citet{moor14,moor15} have produced a comparative observational analysis of the luminosity functions and statistical properties of galaxies present in voids with those present in walls (higher density environment) using data from SDSS and ALFALFA surveys. Besides, they provided empirical fits to the luminosity function of void galaxies.
Further, \cite{moor16} provided a detailed observational study of star formation properties of void galaxies using SDSS data and found void galaxies to harbour a higher star formation rate than galaxies in denser regions like walls. On the other hand, to understand the halo mass function of galaxies in voids, \cite{furl06} devised a semi-analytical expression for the same using the excursion set theory \citep{sheth01}, only replacing the background overdensity term by a similar underdensity to account for the void phenomenon. The problem until this decade had been testing such a model robustly against scarce observations of void galaxies. Now that we have better observations from various surveys, we can put better constraints on void mass functions and properties of void galaxies in general. In our work, we use a similar form of the halo mass function as in \cite{furl06} and a well established semi-analytical star formation model for a single isolated galaxy that correctly describes the evolution of normal galaxies \citep{samui14}, to understand the observations of luminosity functions in large samples of void galaxies.

Apart from the optical measurements, in the recent past, \cite{nero10} reported a lower limit on the  magnetic field, $B\gtrsim$ $10^{-16}$~G, in voids, with coherence length $\lambda_B\gtrsim10^{13}$ cm, through the non-detection of GeV radiation from TeV blazars in Fermi-LAT data. This measurement has opened up several possibilities for the origin and the evolution of the magnetic fields \citep{rbeck01,beckwie13,subr16}. 
A number of explanations that are consistent with the bounds on the void magnetic field strength
and the coherence length (i.e. $B\gtrsim10^{-16}$ G, $\lambda_B\gtrsim10^{13}$ cm) exist,
mainly since the coherence length is very poorly constrained \cite[e.g.,][]{durrer13}. This magnetic field could be primordial,
generated through inflation \citep{turner88}, where they may have arbitrarily large $\lambda_B$. However, numerous models
of inflation exist and therefore there is little predictive power. In contrast, fields generated in electro-weak or QCD phase
transitions in the Early Universe may produce seed magnetic fields $\lesssim10^{-10}$ G with $\lambda_B\lesssim1$ kpc \citep{banerjee04,ichiki06}.
On the other hand, multiple astrophysical explanations of void magnetic field also exist. The Biermann battery effect or Weibel instabilities
in cosmic shocks or forming structures \citep[e.g.,][]{gnedin00,medve07}, for example, can create seed magnetic fields for voids.
However, since voids do not have enough baryonic structure, it is unclear to what extent the Biermann battery can produce seed fields directly,
although studies have found voids very likely to have internal cosmological shocks \citep[e.g.,][]{martin17}. The more likely event is the
battery effect operating in star forming regions to produce seed magnetic fields which can then be amplified and sustained by a galactic dynamo.
Also, \cite{miniati11} have proposed that resistive return currents from cosmic ray propagation in the IGM can produce magnetic fields, $\simeq10^{-16}$G at high redshifts (epoch of reoinization) when the temperatures are low, $\lesssim10^4$ K. However, they used galaxy number density
as measured from Lyman break galaxies as opposed to that in voids.

The other astrophysical possibility is that, the magnetic fields from the galaxies are transported by outflows. Indeed, it has been argued that
the outflows play an important role in polluting the inter galactic medium by metals or magnetic fields in the cluster
environment \citep{madau01,furl03,scan05,bert06,samui08}. In contrast, for voids, \cite{beck13} showed that a low void magnetic field strength could be qualitatively explained if outflows from void galaxies and AGN transport a fraction of magnetic energy stored in those galaxies. However, they neither attempted any detailed quantitative study,
nor their model was constrained from other observations. Here, we use a semi-analytical galactic outflow model devised by \cite{samui18}
and the constrained galaxy formation model that describe the luminosity functions and associated properties of void galaxies to see how much
of the void region can be magnetised
with such outflows that can drag the magnetic field present in void galaxies to the void IGM. In other words, we first tune our star formation/galaxy formation model in voids
such that it reproduces observed luminosity functions of void galaxies and associated properties such as color. Then we use this
galaxy formation model along with a well established outflow model driven by thermal and non-thermal cosmic ray pressures to explore how far
and to what extent galactic magnetic fields can be put in the void region by the outflows originating from the void galaxies themselves
to investigate the possible astrophysical origin of the void magnetic fields.

The paper is organised as follows. In section~2, we describe our models of star formation in galaxies and associated outflows driven by the thermal gas and cosmic rays; in section~3, we constrain the star formation model by fitting the observed luminosity functions of void galaxies; in section~4, we show the resulting magnetisation of voids as obtained from our models. Finally, in section~5, we discuss our results and present conclusions. Throughout this work, we use cosmological parameters as obtained by recent Planck observations \citep{2016A&A...594A..13P}
i.e. $\Omega_{m} = 0.31$, $\Omega_{\Lambda} = 0.69$, $\Omega_{b} = 0.047$ and $h = 0.68$.

\section{Model and Methodology}

\subsection{Halo mass function in voids}
As a parallel to the spherical collapse model, underdense regions can also be described by a simple spherical top-hat form, in an infinite field
with a mean density. These under-dense regions expand and at a certain time various underdensities meet each other resulting in an evacuation
of the centre of the region. At this time, which is referred to as `shell-crossing', the initial model for spherical expansion
does not work anymore. Later on, these voids continue to expand self-similarly \citep[i.e.][]{fill84}.
\citet{furl06} used it to derive a halo mass function for void galaxies by replacing the linearized overdensity of the spherical
collapsed model by a specified value of linearized underdensity.  They showed that at the shell-crossing, linearized underdensity
of the region equals $\delta_v=-2.8$ corresponding to a true underdensity of $\delta=-0.81$. We use it to calculate the formation rate of
dark matter haloes inside a underdense region and is given by \citep{furl06,coor02}
\begin{eqnarray}
n(M; M_v)&=&\sqrt{\frac{2}{\pi}}\frac{\bar{\rho}}{M^2}\bigg\lvert\frac{d\, \ln\,\sigma}{d\, \ln\, M}\bigg\rvert \frac{\sigma^2(\delta_c^L-\delta_v)(1+\delta)}{(\sigma^2-\sigma^2_v)^{3/2}} \nonumber \\ &&~~~~~~~~~~~~~~\times \exp\left[-\frac{(\delta_c^L-\delta_v)^2}{2(\sigma^2-\sigma_v^2)}\right].
\label{eqn_mf}
\end{eqnarray}
Here, $n(M; M_v)$ is the number density of haloes with mass $M$ inside a void region with linearized underdensity $\delta_v$ and mass $M_v$.
Further, $\bar{\rho}$ represents the mean matter density and $\delta^L_c$ provides the critical linearized underdensity required for spherical collapse. Also, $\sigma$ and $\sigma_v$ are the variances in the smoothed density field at the scale of halo and void radius respectively. Note that similar formalism has been used to find the number of satellite haloes inside a bigger halo \citep{2013MNRAS.429.2333J}.

\subsection{Star formation in a galaxy}

Although the formation and evolution of galaxies with their associated star formation histories have been studied comprehensively by numerical simulations as well as semi-analytic models, it is mostly in the region of high density environments like filaments or clusters. It is relatively unknown how galaxy formation/evolution and their star formation work in the context of cosmic voids, which are very low-density environments, with relatively much fewer observations of galaxies. To study them, we consider a star formation model of an individual galaxy that is regulated by the supernova explosion driven outflows from that galaxy.
In \citet{samui14}, this model of star formation was developed for galaxies in the high density universe and we briefly describe it below.
Note that this model was quite successful in explaining the luminosity functions of Lyman Break galaxies in the redshift range $8\ge z \ge 1.5$,
the stellar mass function of haloes with mass from $10^7$~M$_\odot$ to $10^{13}$~M$_\odot$, the 3D correlation between star formation
rate, gas metallicity and stellar mass. Thus we adopt the same for the present work.

We assume that the instantaneous star formation rate at a given time is proportional to the amount of cold gas present in the galaxy.
When a dark matter halo virialises,
the baryonic gas also follows the same and is heated up at the virial temperature. In order to form the stars, the baryonic gas needs to
cool and collapse at the centre of the halo. The cooling rate is govern by the
radiative cooling process and the time scale depends on the composition of the gas. However,
given the condition as seen in haloes, it is generally less than the dynamical time scale of the halo. Thus
we assume that the availability of cold gas at the centre of the galaxy increases due to radiative
cooling with a time scale of the order of the dynamical time scale and is given by \citep{samui14}
\begin{equation}
\frac{dM_g}{dt}=\bigg(\frac{M_b}{\tau}\bigg)e^{-t/\tau},
\label{st4}
\end{equation}
where $M_g$ is the gas mass and $M_b$ equalling $\frac{\Omega_b}{\Omega_m}M$ is the total baryonic mass in the halo.
Further, $\tau$ is the dynamical time of the halo \citep[see][for the detailed expression of it]{bark01}.
After some given time, a certain amount of gas
would be trapped inside stars. On the other hand, the massive stars, which are short-lived, would explode as supernovae after $\sim$ $10^7$ years.
These supernovae have the ability to drive this cold gas out of the galaxy in the form of galactic wind. It is assumed that the amount of the gas in the form of wind would be proportional to the instantaneous star formation rate (neglecting the time delay between the formation of stars and the corresponding supernovae), and is given by
\begin{equation}
\dot{M}_w=\eta_w\dot{M}_{\star},
\label{st5}
\end{equation}
where $M_w$ and $M_{\star}$ are the wind mass and star mass, respectively. The proportionality constant $\eta_w$ varies as a negative index of the circular velocity of the galaxy $v_c$, where the index depends on the mechanism that drives the outflows \citep[for details, see][]{samui08,samui14}.

Finally, there are three factors that will govern the total gas available and hence the star formation rate: (i) gas in stars, (ii) outflowing gas and (iii) the baryonic mass accreted. Therefore, one can write the star formation rate as
\begin{equation}
\frac{dM_{\star}}{dt}=f_t\bigg[\frac{f_{\star}M_g-M_{\star}-M_w}{\tau}\bigg],
\label{st6}
\end{equation}
where $f_t$ is a proportionality constant governing the duration of star formation. Taking the time derivative of Eqn.~\ref{st6} and putting values of $\dot{M}_g$ and $\dot{M}_w$ respectively from Eqns~\ref{st4} and \ref{st5}, one can get
\begin{eqnarray}
\frac{d^2M_{\star}}{dt^2}&=&\frac{f_t}{\tau}\bigg[f_{\star}\frac{dM_g}{dt}-\frac{dM_{\star}}{dt}-\frac{dM_w}{dt}\bigg] \nonumber \\
  &=&\frac{f_t}{\tau}\bigg[f_{\star}\frac{M_b}{\tau} e^{-t/\tau}-\frac{dM_{\star}}{dt}-\eta_w\frac{dM_{\star}}{dt}\bigg].
\label{st7}
\end{eqnarray}
Now if we integrate this equation with boundary conditions that $\frac{dM_{\star}}{dt}=0$ and $M_{\star}=0$ when $t=0$, we have
\begin{equation}
\frac{dM_{\star}}{dt}=\frac{M_bf_{\star}f_t}{\tau[f_t(1+\eta_w)-1]}\bigg[e^{-t/\tau}-e^{-f_t(1+\eta_w)t/\tau}\bigg].
\label{st8}
\end{equation}
Eqn.~\ref{st8} provides the analytical form of the star formation rate in a galaxy at age $t$ in the presence of supernova
feedback.

\subsection{Luminosity Function}
\label{sec_lf}
The above mentioned star formation rate is converted to the luminosity as follows.
We use the STARBURST99 code \citep{leith99} to obtain the time evolution of luminosity, $l_{\lambda}(t)$, at any particular wavelength band ($\lambda$)
for a single burst of star formation given an initial mass function for stars formed. Here we consider the median of SDSS $r$ and $u$ bands, which are 623~nm and 354~nm respectively, redshifted to $z=0.1$ \citep[see][]{moor15}. Further, we assume Salpeter initial mass function in the mass range
$0.1-100$~M$_\odot$.
The luminosity, $l_{\lambda}(t)$, is in turn convolved with the star formation rate (i.e. Eqn.~\ref{st8}) of the corresponding galaxy to obtain the luminosity as a function of galaxy age. Here star formation is continuous and lasting for a few dynamical time. The luminosity of a galaxy of age $T$ and mass $M$ is hence calculated using
\begin{equation}
L_{\lambda}(M,z,z_c)\equiv L_{\lambda}(M,T)=\int_{T}^{0}\dot{M}_{*}(M,T,T-\tau)l_\lambda(\tau)d\tau,
\end{equation}
where the age of the galaxy formed at $z_c$ and observed at $z$ is given by $T(z,z_c)=t(z)-t(z_c)$, $t(z)$ being the age of the universe at $z$.
Also, we have used a correcting factor, $1/\eta$, to adjust for the dust absorption, which also depends on the wavelength.
The luminosity can be converted to a standard absolute magnitude ($M_{AB}$) in the AB system.

Finally, the luminosity function, $\Phi(M_{AB},z,\lambda)$,
at redshift $z$ and
at a wavelength $\lambda$ can be written as
\begin{eqnarray}
&&\Phi(M_{AB},z,\lambda)dM_{AB} \nonumber \\
&&~~~~=\int\limits^{\infty}_{z}dz_c\int\limits_{M_{low}}^{\infty}N(M,z,z_c)\frac{dM}{dL_{\lambda}}\frac{dL_{\lambda}}{dM_{AB}}dM_{AB}.
\end{eqnarray}
Here $N(M,z,z_c)$ is the number of the dark matter haloes per unit volume between mass $M$ and $M+dM$ which collapsed between redshift $z_c$ and $z_c+dz_c$.
As noted earlier, we use the form of galaxy mass function in voids given by \citet{furl06}.
Further, $M_{low}$ is decided as follows.
A galaxy forming in a neutral region can cool with the help of atomic cooling if the virial temperature of the dark matter halo is more than $10^4$~K \citep{bark01}. Thus $M_{low}$ corresponds to the halo mass having virial temperature $10^4~$K.

In the ionized universe, an increase of Jeans mass will allow a galaxy to host star formation only when its virial velocity goes above 35 km/s
\citep{1996ApJ...465..608T}; because of this, we assume a total suppression of star formation for velocities less than that (i.e. $M_{low}$). Simulations also show that galaxies with a virial velocity more than 95~km/s are not affected by the radiative feedback at all \citep{1996ApJ...465..608T}. However, it was shown in \citet{samui18a} that a higher virial velocity cut-off of 110~km/s provides a better fit to the observed luminosity functions of high redshift Lyman break galaxies, and hence we use it here.
 For intermediate velocities (intermediate mass range), we use a linear fit from 1 to 0 \citep{bromm02,benson02,dijk04},
i.e., a linear interpolation between 35~km/s and 110~km/s
such that at $\sim  72$~km/s, there is 50\% suppression, complete suppression at and below 35~km/s
and zero or no suppression at and above 110~km/s.
 Finally, we also assume a decrease in star formation in high mass haloes due to possible AGN feedback, by a factor of $[1+(M/10^{12}M_{\odot})^3]^{-1}$ \citep{samui14}.

\subsection{Galactic Outflows}
We use \cite{samui18} model of outflows that are driven by thermal gas and non-thermal cosmic rays (CRs) which provide an extra pressure in the outflow dynamics.
The detailed modeling can be found in \citet{samui08} and \citet{samui18}.
We assume a spherically symmetric thin shell model of outflows from galaxies. Initial supernovae (SNe) explosions result
in the creation of so-called ‘supernova bubble’ of hot gas that expands as subsequent SNe explosions produce a free wind that converts the kinetic energy into thermal energy. This free wind consists of thermal as well as high-energy CR particles which are produced at the SNe terminal shocks. The shock due to the free wind is called as `inner shock' of the outflow.
At the inner shock, diffusive shock acceleration mechanism can accelerate the CR particles with efficiency as high as $\sim$ 50\% \citep{kangj02,kangj03,kangj05}.
The thermal pressure, as well as the cosmic ray pressure in the hot gas bubble, creates and drives an outer shock that sweeps IGM materials. This swept up material is confined to a thin shell region and is separated from the shocked free wind by a contact discontinuity.

In presence of cosmic rays and hot gas, the time evolution of the outer shock position of a spherically symmetric outflow is governed by the following equations \citep{samui18},
\begin{equation}
\frac{d^2R_s}{dt^2}=\frac{4\pi R_s^2[P_b+P_c-P_0]}{m_s(R_s)}-\frac{{\dot{m}}_s(R_s)\Big[{\dot{R}}_s-v_0(R_s)\Big]}{m_s(R_s)}-\frac{GM_s}{R_s^2}
\label{eqn_r}
\end{equation}
and
\begin{equation}
\frac{d}{dt}[m_s(R_s)]=\epsilon 4\pi R_s^2\rho_B(R_s)\Big[\dot{R_s}-v_0(R_s)\Big],
\label{eqn_dm_dt}
\end{equation}
where $R_s$ is the position of the outer shock and $m_s$ is the mass of swept up gas contained in the thin shell. Further, $P_b$ and $P_c$ are the thermal and cosmic ray pressures in the hot bubble. In addition, $P_0$ is the IGM pressure assuming a temperature of $10^4$~K and $v_0$ is the ambient velocity field \citep{furl01}.

In the first equation (i.e. Eqn.~\ref{eqn_r}), the right hand side consists of three terms. The first term represents the net pressure driving the outer shock; the second term is simply mass accumulation/loading from the medium outside (with baryonic density $\rho_B$) leading to deceleration (hence the negative sign).
And as evident, the final term in the first equation represents the gravitational attraction due to dark matter
\citep[a NFW profile is assumed with concentration parameter $c = 4.8$,][]{navar97,madau01}, as well as baryonic matter within radius $R_s$ with total mass $M_s$.
The second equation (i.e. Eqn.~\ref{eqn_dm_dt}) represents the dynamic evolution of the gas mass in the thin shell, where it is assumed a fraction, $\epsilon$ of the swept-up mass remains in the shell, and the remaining is mixed with the hot bubble due to evaporation and/or fragmentation.

The thermal energy, $E_b$ and the cosmic ray energy, $E_c$ in the hot bubble
evolve as
\begin{equation}
\frac{\mathrm{d} E_{\mathrm{b}}}{\mathrm{d} t}=L(t)-4 \pi\left(R_{\mathrm{s}}^{2} \dot{R}_{\mathrm{s}}-R_{1}^{2} \dot{R}_{1}\right) P_{\mathrm{b}}-\Lambda(t, T, Z),
\label{eq13}
\end{equation}
and
\begin{equation}
\frac{\mathrm{d} E_{c}}{\mathrm{d} t}=L_{c}(t)-4 \pi\left[R_{s}^{2} \dot{R}_{s}-R_{1}^{2} \dot{R}_{1}\right] P_{c},
\label{eq14}
\end{equation}
where $R_1$ is the position of inner shock. The thermal and cosmic ray pressures are related to the thermal and cosmic ray energies by usual thermodynamic relations with adiabatic indices of 5/3 and 4/3 respectively. Further, $L(t)$ and $L_c(t)$ are the total thermal energy and the cosmic ray energy injected in the hot bubble per unit time. They are related to the star formation rate ($\frac{dM_*}{dt}$) of the galaxy as
\begin{equation}
L(t)=10^{51}\,{\rm ergs}\times\epsilon_w\nu\frac{dM_*}{dt}
\label{eq_L}
\end{equation}
and
\begin{equation}
L_c(t)=10^{51}\,{\rm ergs}\times\epsilon_{cr}\nu\frac{dM_*}{dt}.
\label{eq_Lc}
\end{equation}
Here we assume that $\nu$ number of SNe are formed per unit mass of star formation and on an average $10^{51}$~ergs of energy are produced per SNe.
Further, a fraction, $\epsilon_w$, of the total SNe energy is transferred to the hot bubble as thermal energy and another fraction, $\epsilon_{cr}$ is channeled into the cosmic ray energy.
Hence the total efficiency in converting SNe energy
to drive the outflow is simply $\epsilon_w$+$\epsilon_{cr}$. For our calculations we assume $\epsilon_w$=$\epsilon_{cr}$= 0.15;
the efficiency is set from results of numerical simulations \citep{mori02}.

The 2nd terms in the right hand side of Eqns.~\ref{eq13} and \ref{eq14} represent the work done due to expansion. Finally, $\Lambda$ in Eqn.~\ref{eq13} describes the thermal cooling of the hot gas and has been calculated assuming a uniform density of the hot bubble region. We consider cooling due to Compton drag against the cosmic
microwave background radiation, bremsstrahlung and recombination line cooling.
Note that the thermal cooling is very important and can even stop the outflow in low mass galaxies to originate. Only in the presence of cosmic rays, outflows can be generated in such small galaxies which (as we will show) play a major role in volume filling of the outflowing material and magnetic field in voids. Further, it was shown in \citet{samui18} that in high mass galaxies with mass above $10^{12}$~M$_\odot$, the presence of cosmic rays has a negligible effect on the outflow dynamics. These are consistent with simulations \citep[i.e.][]{2012MNRAS.423.2374U,2018MNRAS.477..531F} making our outflow models more robust.

All of the above equations are solved numerically as described in detail in \cite{samui18} and \citet{samui08}. We follow the dynamics of the outflow until the outflow peculiar velocity decreases to the local sound speed in the IGM and cannot drive itself hence further. The shock disappears and the outflowing material mixes with the IGM to finally merge with the Hubble flow.

The global impact of these outflows is relevant in our case as we want to investigate the degree of percolation of these outflows from the galaxies
into the void region and the extent to which one can explain the observed void magnetic fields,
through studying the pertinent volume filling factor and the magnetic field.
The volume filling factor $F(z)=1-\exp[-Q(z)]$ indicates the volume of the IGM affected by the outflow in the case where there is no source clustering and they are randomly distributed, where the porosity $Q$ is given as
\begin{equation}
Q(z)=\int^{\infty}_{M_{low}}dM\int^{\infty}_{z}dz'~N(M,z,z')\frac{4\pi}{3}[R_s(1+z)]^3.
\end{equation}
Here $N(M,z,z')$ is the corresponding halo formation rate, and $R_s$ is the size of the outflow bubble
created by that galaxy at $z$ obtained by solving Eqn.~\ref{eqn_r} upto the time $t(z')-t(z)$.
The lower mass limit in the above equation is calculated from the physical conditions that are required to host star formation as already described
in section~\ref{sec_lf}.

Note that in our thin shell model of outflow, the outflowing matter from the galaxy is accumulated at the contact discontinuity. However, it is likely to get mixed due to instabilities like the Rayleigh-Taylor (RT) instability. In our previous work i.e. \citet{samui08}, the detail investigation of these instabilities was carried out and it was shown that within the time scale of outflows the RT instability is likely to mix the material at contact discontinuity. However, this does not change the dynamics of the outer shock. Further, it was shown \citep[Fig.~B1 of][]{samui08} that the contact discontinuity is likely to be very close to the outer shock. Thus considering the outer shock radius of outflow to calculate the total volume filling by the material/magnetic field won't make much of a difference.

\subsection{Magnetic field dynamics}

The main goal of this work is to understand the magnetisation of the void IGM through outflows from the galaxies residing inside the void itself.
The magnetic fields are assumed to be generated in the ISM of high redshift proto-galaxies through amplification of small seed magnetic fields. We assume an initial microgauss magnetic field in the galaxy and allow the outflow and the magnetic field to evolve separately. This excludes dynamical effects of the field on the outflow. Since the ionized plasma in the outflow is coupled with the magnetic field, the thermal and cosmic rays driven outflow, is likely to carry the magnetic fields out of the galaxy in addition of carrying the metals. For calculating the strength of the magnetic field that can be deposited in the void IGM via this process, we follow the prescription of \cite{bert06} where they considered magnetising the general IGM by thermally driven winds.

Since we are interested to know the minimum field that can be put by these outflows, we ignore any possible amplification of the magnetic field due to shear flow and turbulence of the outflowing material.
In such a case, the magnetic field dynamics simply involves the insertion of the magnetic field into the outflow material,
which carries the field out of the galaxy resulting in a decreasing magnetic field over time.
This obeys the magnetic flux freezing condition throughout. Thus the magnetic energy ($E_B$) evolution is governed by
\citep{bert06}
\begin{equation}
\frac{dE_B}{dt}={\dot{E}}_{B_{in}}-\frac{1}{3}\frac{{\dot{V}}_w}{V_w}E_B,
\end{equation}
where $V_w$ is the total volume of the wind bubble and ${\dot{E}}_{B_{in}}$ is the magnetic energy injection rate in the bubble from the given galaxy
from the star forming region.
The injection rate ${\dot{E}}_{B_{in}}$ is taken as
\begin{equation}
\dot{E}_{B_{in}}=\epsilon_{B_{in}}\frac{{\dot{M}}_w}{\bar{\rho}_{in}},
\end{equation}

where the average density of the injected free wind material at the inner shock is given by
\begin{equation}
\bar{\rho}_{i n}=\frac{\dot{M}_{w}}{4 \pi R_{1}^{2} v_{w}}.
\end{equation}
Here, $v_w$ is the speed of free wind material and obtained from $L(t)=\dot{M}_w(t)v_w^2/2$.

Moreover, the injected magnetic energy density is given by
\begin{equation}
\epsilon_{B_{in}}=\frac{B_0^2}{8\pi}\bigg[\frac{\bar{\rho}_{in}}{\bar{\rho}_{ISM}}\bigg]^{4/3}
\label{eqn_B}
\end{equation}
where $B_0$ is the magnetic field inside the star forming galaxy. We have taken $B_0=10~\mu$G as has been observed in local galaxies \citep{rbeck15}.
Further, we have assumed 1000 times the average baryonic density of the halo as $\bar{\rho}_{ISM}$ (see \citealp{samui18} for details).

\section{Luminosity function of voids}
\begin{figure}
\centerline{
\epsfig{figure=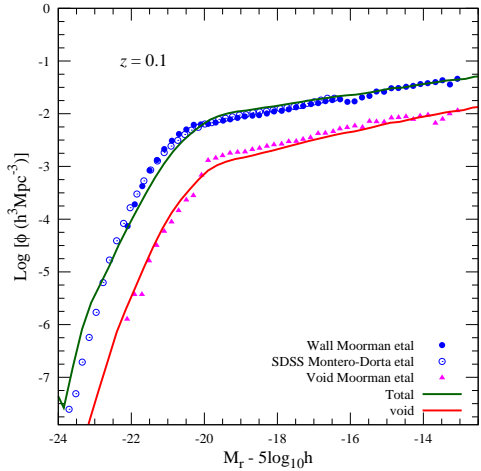,width=6.5cm,angle=-90.}%
}
\caption[]{$r$-band luminosity functions at $z=0.1$ as predicted by our models for void (solid red curve) and wall galaxies
(solid dark green curve). The observational data are taken from \cite{moor15} (filled circles and triangles)
and \cite{monter09} (open circles).}
\label{fig_LFr}
\end{figure}

We first concentrate on the luminosity functions of void galaxies as predicted by our model.
As already mentioned, \cite{moor15} has measured void
luminosity functions at $z=0.1$ from the SDSS DR7 catalog in the $r$ band.
Further, they have compared the luminosity function of void galaxies to that of wall galaxies, and compared their colors as well.
Note that the median effective radius of
their voids is $\sim 17~h^{-1}$~Mpc. This has been used to calculate the halo mass function in voids in our model.
In Fig.~\ref{fig_LFr}, we show the $r$-band luminosity function for void galaxies as obtaibed from
our model with the solid red curve
along with the data points taken from \cite{moor15} (magenta filled triangles). As already mentioned we adjust the dust reddening
correction factor to match our result with the observation. For comparison, we also show $r$-band
luminosity functions for wall galaxies, both our model prediction (solid dark green curve) and observational data (open and
filled circles). For wall galaxy luminosity function we use the Press-Schechter formalism with \cite{sasaki94} prescription
for finding the formation rate of haloes. At low redshifts, galaxy mergers become very important. Thus the survival probability
of a galaxy must be included while obtaining luminosity functions and that's why we prefer to use
the Sasaki formalism over naive derivative to calculate the formation rate of halos (see \citealp{samui09} for details).
The Sasaki formalism \citep{sasaki94} takes acount of the merger of galaxies by considering the survival
probability of a galaxy at a later epoch with the assumption that the efficiency of destruction rate
has no characteristic mass scale.
It is clear from the figure that our model predictions match well with the
observational data both for void galaxies and wall galaxies. The low end slope and the extent of the luminosity
function are well described by our SNe feedback regulated star formation model. Further, the high end of the luminosity
function is also well explained by our AGN feedback model both for void galaxies and wall galaxies. Thus
the entire luminosity function from M$_{\rm AB} = -24$ upto M$_{\rm AB} = -13$ spanning almost five orders in
luminosity can be fitted with our model. Therefore, we conclude that SNe feedback is operational
not only in field galaxies but also in galaxies residing in the extreme void regions.

\begin{figure}
\centerline{
\epsfig{figure=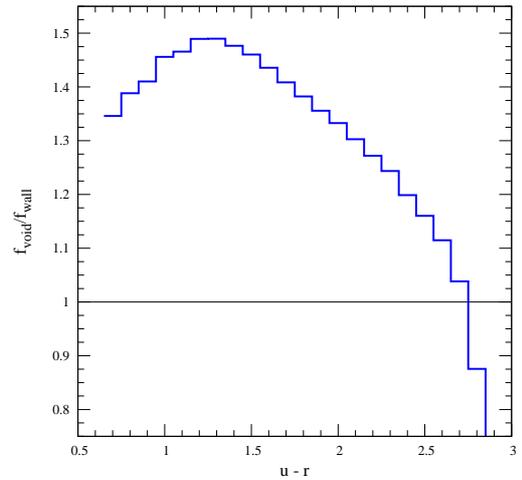,width=6.5cm,angle=-90.}%
}
\caption[]{The ratio of fraction of void galaxies to fraction of wall galaxies as a function of $u-r$ color
as predicted by our model.}
\label{fig_u-r}
\end{figure}

In order to further investigate the validity of our model, we explore the $u-r$ color of galaxies.
Note that \cite{moor15} showed that void galaxies are comparatively bluer compared to the
wall galaxies. In Fig~\ref{fig_u-r}, we have plotted the fractional ratio of the void to wall galaxies
as a function of their color.
Clearly, our models also predict that the void galaxies are bluer compared to the wall galaxies
as obtained in the observation.

\begin{figure}
\centerline{
\epsfig{figure=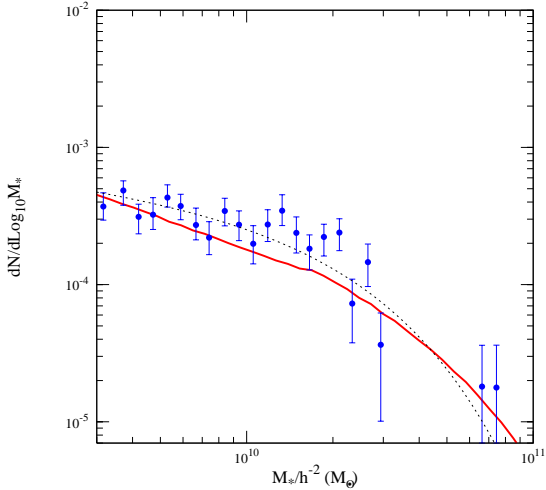,width=6.5cm,angle=-90.}%
}
\caption[]{The stellar mass function of void galaxies as predicted from our model (solid red curve).
The data points are taken from \citet{2015MNRAS.451.3249A} along with the Schechter function fit shown by the dotted curve.
}
\label{fig_smf}
\end{figure}

We also calculate the stellar mass function of the void galaxies and compare it with the available observations.
In Fig.~\ref{fig_smf} we have shown the predicted stellar mass for void galaxies at $z=0$ with the solid red curve along with
the observational data taken from \citet{2015MNRAS.451.3249A}. The dotted curve is the Schechter function fit as given in \citet{2015MNRAS.451.3249A}.
It is clear from the figure that the predicted staller mass function of void galaxies matches quite well
with the observations making our model more robust.

Thus we conclude that our star formation and galaxy formation models are very successful in explaining the
observed properties of void galaxies. With this constrained model, we now explore the possible magnetization
of the void region by the outflows from these void galaxies.

\section{Outflows and magnetisation}
\begin{figure}
\centerline{
\epsfig{figure=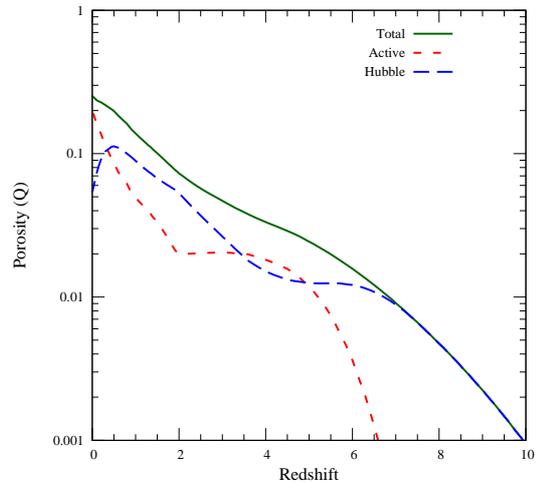,width=6.5cm,angle=-90.}%
}
\caption[]{Porosity $Q$ of the outflows as a function of redshift. We show the total $Q$ by the solid
dark green curve. The blue long dashed and red short dashed curve represent the contribution from the Hubble frozen
and active outflows respectively.}
\label{fig_Q}
\end{figure}

In this section, we investigate how far the outflows from the void galaxies can magnetise the void IGM.
We first concentrate on the volume filling factor of outflows. In Fig.~\ref{fig_Q}, we show the porosity
of outflows as a function of the redshift for a void of size $17 h^{-1}$~Mpc.
The total $Q$ has also been split into two parts: one
by outflows that are active, and are still producing shocks (the red short-dashed curve) while the other by outflows that are just evolving with the Hubble flow (the blue long-dashed curve). It is clear from the figure that almost 25\% of the volume of the void can be polluted with materials from outflows. Out of which, 20\% contribution comes from the active
outflows whereas rest by outflows moving with the Hubble flow. It is important to note here that at some point ($z\sim 1$) the volume filling factor by the dead outflows is $\sim 10\%$ which then decreases. This is due to the way we treat the outflows of merging galaxies. We assume that if two galaxies merge together, then their outflows get destroyed and a fresh outflow starts. This leads to a decrease of the volume filling factor by the dead outflows.
If we add these the total volume factor or $Q$ would be more than 0.3. See \cite{samui18} for more detail discussion of this. Thus we conclude that the outflows from void galaxies themselves can percolate $\sim 30\%$ volume of the voids. Since these winds are driven by cosmic rays they will also carry the magnetic field with them. Therefore, $\sim 30\%$ of the void can be filled with magnetic field taken from the galaxies by the outflows.

\begin{figure}
\centerline{
\epsfig{figure=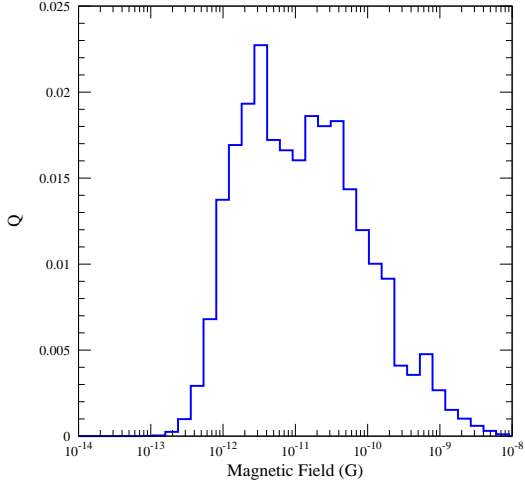,width=6.5cm,angle=-90.}%
}
\caption[]{The porosity as a function of magnetic field strength in voids as obtained by our model at $z=0$.}
\label{fig_B}
\end{figure}

\begin{figure}
\centerline{
\epsfig{figure=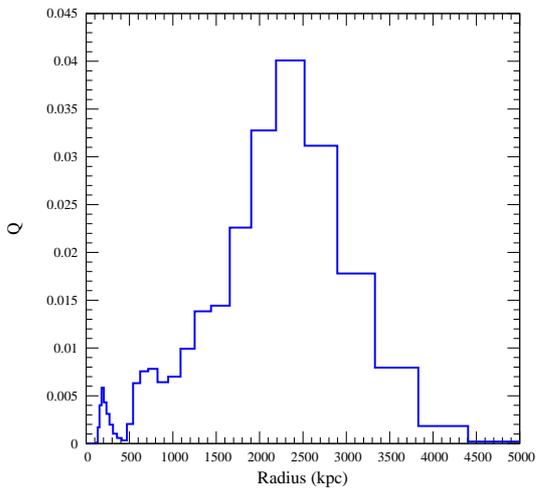,width=6.5cm,angle=-90.}%
}
\caption[]{Porosity distribution of outflows with different outflow radius at $z=0$.}
\label{fig_R}
\end{figure}

\begin{figure}
\centerline{
\epsfig{figure=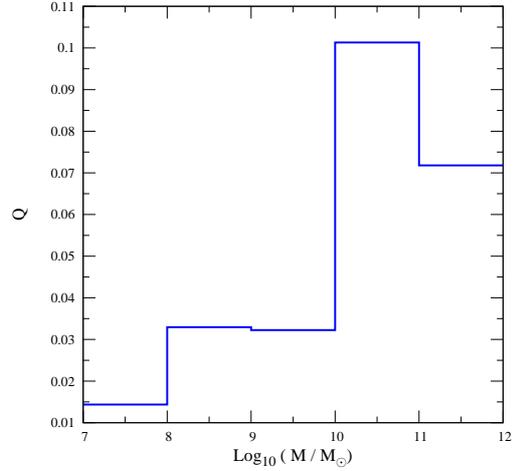,width=6.5cm,angle=-90.}%
}
\caption[]{Contribution of different halo masses to the porosity, $Q$.}
\label{fig_M}
\end{figure}

To investigate in more details, we show in Fig.~\ref{fig_B}
the porosity $Q$ as a function of magnetic field strength as of today (i.e. $z=0$). It is evident from the figure that most
of the regions is filled with the magnetic field strength of $10^{-12} - 10^{-10}~$G. This is higher than the
lower limit of the magnetic field strength as inferred from the blazar observations  \citep{nero10}. Thus we can say
that the magnetic field as observed in the voids are of astrophysical origin, originating in the void
galaxies and transported to the void IGM by the outflows driven by the hot thermal gas and cosmic rays.
Further, in Fig.~\ref{fig_R}, we show the size distribution of individual outflows. This reveals that
most of the bubble sizes are more than a megapersec. This is again consistent with the coherent
length scale of the observed magnetic field  \cite{nero10}. Thus these outflows clearly explain all available characteristics
of the void magnetic field and establish the astrophysical origin of the void magnetic field.
Further, in Fig.~\ref{fig_M} we show the contribution to $Q$ by different halo mass range. This shows
that outflows originating in halo mass range $10^{10}-10^{12}$~M$_\odot$ contribute most in volume filling the
voids with magnetic field.

Note that the total efficiency of the wind that we have assumed is $\epsilon_{cr}+\epsilon_w = 0.3$.
It is the most important free parameter in our model that decides the volume filling factor and hence the magnetization. Note that parameters related to the star formation model have been fixed by fitting the observed luminosity function.
If the efficiency is higher then the volume filling would be more with less average magnetic field strength due to the flux freezing. On the other
hand if the efficiency is lower the field strength would be higher with less volume filled.
It is important to note that even if outflows do not completely volume fill the voids, one outflow
with megapersec size in the line of sight is enough to explain the spectrum of TeV blazars that
established the existence of void magnetic field  \cite{nero10}. Randomly distributed sources with 1/4th volume filled are enough to have a high probability of crossing one outflowing region by the line of sight causing the deflection of GeV photons in the blazar spectrum (see \ref{app} for a detailed calculation).

Further, the amount of magnetisation in the voids is directly related to the magnetic energy input from the outflow originating galaxy, which is decided by $B_0$ (see Eqn.~\ref{eqn_B}). We have used $B_0=10~\mu$G that resulted the void magnetic field of order $10^{-10}-10^{-12}$~G. Thus even a much much lower magnetic filed input from the galaxy would be consistent with the observed lower limit of $B\gtrsim 10^{-16}~$G in voids.

\section{Discussion and conclusions}
In this work, we investigate a possible astrophysical origin of the void magnetic fields as detected by the observations of blazars.
We estimate the amount and extent of the magnetic field that can be put in the void IGM. The magnetic fields are likely to be present in the void galaxies
themselves and can be transported by outflows from these galaxies in the presence of charged cosmic rays.
To explore that we use excursion set theorem to find the number of halos that can form in an underdense environment i.e. the void.
The star formation rate in a single void galaxy is calculated
taking account of feedback due to the supernova driven outflow. The star formation rate is converted to the luminosity using a population synthesis
code STARBURST99. With these, we calculate the $r$-band luminosity functions of void galaxies and show that our model provides an excellent
description of the observed luminosity functions. Further, our model also predicts that the void galaxies are bluer compared to their
counterparts in the denser environment. This is also consistent with observations.

Using this galaxy formation/evolution model, we then estimate the amount of magnetisation in the void IGM. To do that we consider
an outflow model that assumes both hot thermal gas and non-thermal cosmic rays providing the driving force for the outflow. We show that
such a model can magnetise $\sim 30\%$ of the void IGM by today. Most of the IGM is magnetised with the magnetic field strength of $10^{-12}
-10^{-10}$~G, which is higher than the lower limit of the detected magnetic fields. On an average, most of the outflows have a radius of
megapersec which is comparable to the coherent length scale of the observed magnetic field.
These values depend on the energy efficiency of outflows and the magnetic field input by the void galaxies. We further
argued that even though our model does not volume fill the void IGM and the volume filling factor depends on the efficiency of the outflow, it can
potentially explain the observed missing GeV photons in the TeV blazars.

Thus we conclude that the detected magnetic field in the void could have come from the few galaxies that are there, transported to the void intergalactic medium
by the outflows generated by the supernova, that resulted in a null detection of GeV photons in the
TeV blazars.

However, we cannot strongly dismiss the possibility of primordial or any other sources for the origin of this void magnetic field. This requires future sensitive observations to constrain the field strength and coherent length scale separately.

\section*{Acknowledgments}
SS thanks IUCAA, India for support through its associateship programme. SS also thanks Presidency University, Kolkata, India for providing funds through FRPDF scheme. SS also acknowledges  financial  support  from BRNS through a project grant (sanction no: 57/14/10/2019-BRNS). ARC received support from the DST INSPIRE fellowship.

\appendix
\section{Probability of line-of-sight encountering outflow inside void}
\label{app}
We being by considering a void of size $24$ Mpc which is $\sim17h^{-1}$ Mpc used in our calculations. We populate this void in increasing fractions by outflows of diameter $1$ Mpc. For each fraction of void filled, we compute a probability distribution of the maximum length of a single outflow that a random line of sight will encounter. We plot below in Figure \ref{los} a cumulative probability of a line of sight to have traversed $>X$ Mpc through an outflow.
\begin{figure}[h!]
\centerline{
\epsfig{figure=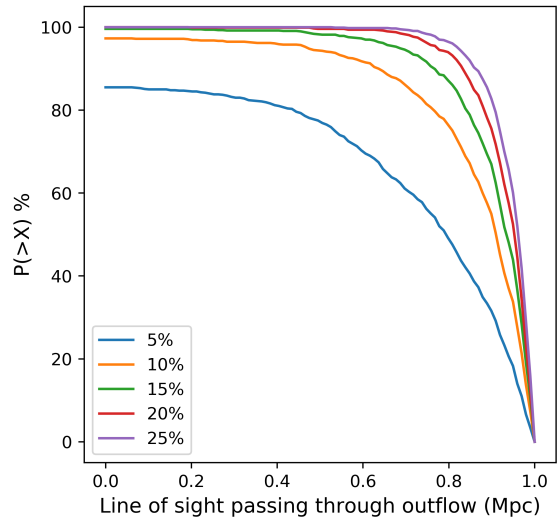,width=8cm,angle=0.}%
}
\caption[]{Probabilities of encountering outflow of more than given size for different void filling fractions.}
\label{los}
\end{figure}

It is clear from the above figure that if only $10\%$ of the void is filled, there is a $\sim$ $80\%$ chance a line of sight will pass through an outflow of size $\gtrapprox$ $0.8$ Mpc. This confirms that outflows from supernovae have a very high likelihood of deflecting GeV photons in the blazar spectrum.

\bibliography{void_lf_mag_r1.bib}

\begin{thebibliography}{67}
\expandafter\ifx\csname natexlab\endcsname\relax\def\natexlab#1{#1}\fi
\providecommand{\url}[1]{\texttt{#1}}
\providecommand{\href}[2]{#2}
\providecommand{\path}[1]{#1}
\providecommand{\DOIprefix}{doi:}
\providecommand{\ArXivprefix}{arXiv:}
\providecommand{\URLprefix}{URL: }
\providecommand{\Pubmedprefix}{pmid:}
\providecommand{\doi}[1]{\href{http://dx.doi.org/#1}{\path{#1}}}
\providecommand{\Pubmed}[1]{\href{pmid:#1}{\path{#1}}}
\providecommand{\bibinfo}[2]{#2}
\ifx\xfnm\relax \def\xfnm[#1]{\unskip,\space#1}\fi
\bibitem[{{Kirshner} et~al.(1981){Kirshner}, {Oemler}, {Schechter}, and
  {Shectman}}]{kirsh81}
\bibinfo{author}{R.~P. {Kirshner}}, \bibinfo{author}{A.~{Oemler}, Jr.},
  \bibinfo{author}{P.~L. {Schechter}}, \bibinfo{author}{S.~A. {Shectman}},
\newblock \bibinfo{title}{{A million cubic megaparsec void in Bootes}},
\newblock \bibinfo{journal}{\apjl} \bibinfo{volume}{248} (\bibinfo{year}{1981})
  \bibinfo{pages}{L57--L60}. \DOIprefix\doi{10.1086/183623}.
\bibitem[{{Kauffmann} and {Fairall}(1991)}]{kauff91}
\bibinfo{author}{G.~{Kauffmann}}, \bibinfo{author}{A.~P. {Fairall}},
\newblock \bibinfo{title}{{Voids in the distribution of galaxies - an
  assessment of their significance and derivation of a void spectrum}},
\newblock \bibinfo{journal}{\mnras} \bibinfo{volume}{248}
  (\bibinfo{year}{1991}) \bibinfo{pages}{313--324}.
  \DOIprefix\doi{10.1093/mnras/248.2.313}.
\bibitem[{{El-Ad} et~al.(1996){El-Ad}, {Piran}, and {da Costa}}]{elad96}
\bibinfo{author}{H.~{El-Ad}}, \bibinfo{author}{T.~{Piran}},
  \bibinfo{author}{L.~N. {da Costa}},
\newblock \bibinfo{title}{{Automated Detection of Voids in Redshift Surveys}},
\newblock \bibinfo{journal}{\apjl} \bibinfo{volume}{462} (\bibinfo{year}{1996})
  \bibinfo{pages}{L13}. \DOIprefix\doi{10.1086/310027}.
  \href{http://arxiv.org/abs/astro-ph/9512070}{{\tt arXiv:astro-ph/9512070}}.
\bibitem[{{El-Ad} and {Piran}(1997)}]{elad97}
\bibinfo{author}{H.~{El-Ad}}, \bibinfo{author}{T.~{Piran}},
\newblock \bibinfo{title}{{Voids in the Large-Scale Structure}},
\newblock \bibinfo{journal}{\apj} \bibinfo{volume}{491} (\bibinfo{year}{1997})
  \bibinfo{pages}{421--435}. \DOIprefix\doi{10.1086/304973}.
  \href{http://arxiv.org/abs/astro-ph/9702135}{{\tt arXiv:astro-ph/9702135}}.
\bibitem[{{Hoyle} et~al.(2002){Hoyle}, {Vogeley}, {Gott}, {Blanton}, {Tegmark},
  {Weinberg}, {Bahcall}, {Brinkmann}, and {York}}]{hoyle02}
\bibinfo{author}{F.~{Hoyle}}, \bibinfo{author}{M.~S. {Vogeley}},
  \bibinfo{author}{J.~R. {Gott}, III}, \bibinfo{author}{M.~{Blanton}},
  \bibinfo{author}{M.~{Tegmark}}, \bibinfo{author}{D.~H. {Weinberg}},
  \bibinfo{author}{N.~{Bahcall}}, \bibinfo{author}{J.~{Brinkmann}},
  \bibinfo{author}{D.~{York}},
\newblock \bibinfo{title}{{Two-dimensional Topology of the Sloan Digital Sky
  Survey}},
\newblock \bibinfo{journal}{\apj} \bibinfo{volume}{580} (\bibinfo{year}{2002})
  \bibinfo{pages}{663--671}. \DOIprefix\doi{10.1086/343734}.
  \href{http://arxiv.org/abs/astro-ph/0206146}{{\tt arXiv:astro-ph/0206146}}.
\bibitem[{{Platen} et~al.(2007){Platen}, {van de Weygaert}, and
  {Jones}}]{platen07}
\bibinfo{author}{E.~{Platen}}, \bibinfo{author}{R.~{van de Weygaert}},
  \bibinfo{author}{B.~J.~T. {Jones}},
\newblock \bibinfo{title}{{A cosmic watershed: the WVF void detection
  technique}},
\newblock \bibinfo{journal}{\mnras} \bibinfo{volume}{380}
  (\bibinfo{year}{2007}) \bibinfo{pages}{551--570}.
  \DOIprefix\doi{10.1111/j.1365-2966.2007.12125.x}.
  \href{http://arxiv.org/abs/0706.2788}{{\tt arXiv:0706.2788}}.
\bibitem[{{Sheth} and {van de Weygaert}(2004)}]{sheth04}
\bibinfo{author}{R.~K. {Sheth}}, \bibinfo{author}{R.~{van de Weygaert}},
\newblock \bibinfo{title}{{A hierarchy of voids: much ado about nothing}},
\newblock \bibinfo{journal}{\mnras} \bibinfo{volume}{350}
  (\bibinfo{year}{2004}) \bibinfo{pages}{517--538}.
  \DOIprefix\doi{10.1111/j.1365-2966.2004.07661.x}.
  \href{http://arxiv.org/abs/astro-ph/0311260}{{\tt arXiv:astro-ph/0311260}}.
\bibitem[{{Patiri} et~al.(2006){Patiri}, {Betancort-Rijo}, and
  {Prada}}]{2006MNRAS.368.1132P}
\bibinfo{author}{S.~G. {Patiri}}, \bibinfo{author}{J.~{Betancort-Rijo}},
  \bibinfo{author}{F.~{Prada}},
\newblock \bibinfo{title}{{On an analytical framework for voids: their
  abundances, density profiles and local mass functions}},
\newblock \bibinfo{journal}{\mnras} \bibinfo{volume}{368}
  (\bibinfo{year}{2006}) \bibinfo{pages}{1132--1144}.
  \DOIprefix\doi{10.1111/j.1365-2966.2006.10202.x}.
  \href{http://arxiv.org/abs/astro-ph/0407513}{{\tt arXiv:astro-ph/0407513}}.
\bibitem[{{Conroy} et~al.(2005){Conroy}, {Coil}, {White}, {Newman}, {Yan},
  {Cooper}, {Gerke}, {Davis}, and {Koo}}]{2005ApJ...635..990C}
\bibinfo{author}{C.~{Conroy}}, \bibinfo{author}{A.~L. {Coil}},
  \bibinfo{author}{M.~{White}}, \bibinfo{author}{J.~A. {Newman}},
  \bibinfo{author}{R.~{Yan}}, \bibinfo{author}{M.~C. {Cooper}},
  \bibinfo{author}{B.~F. {Gerke}}, \bibinfo{author}{M.~{Davis}},
  \bibinfo{author}{D.~C. {Koo}},
\newblock \bibinfo{title}{{The DEEP2 Galaxy Redshift Survey: The Evolution of
  Void Statistics from z \raisebox{-0.5ex}\textasciitilde 1 to z
  \raisebox{-0.5ex}\textasciitilde 0}},
\newblock \bibinfo{journal}{\apj} \bibinfo{volume}{635} (\bibinfo{year}{2005})
  \bibinfo{pages}{990--1005}. \DOIprefix\doi{10.1086/497682}.
  \href{http://arxiv.org/abs/astro-ph/0508250}{{\tt arXiv:astro-ph/0508250}}.
\bibitem[{{Goldberg} et~al.(2005){Goldberg}, {Jones}, {Hoyle}, {Rojas},
  {Vogeley}, and {Blanton}}]{gold05}
\bibinfo{author}{D.~M. {Goldberg}}, \bibinfo{author}{T.~D. {Jones}},
  \bibinfo{author}{F.~{Hoyle}}, \bibinfo{author}{R.~R. {Rojas}},
  \bibinfo{author}{M.~S. {Vogeley}}, \bibinfo{author}{M.~R. {Blanton}},
\newblock \bibinfo{title}{{The Mass Function of Void Galaxies in the Sloan
  Digital Sky Survey Data Release 2}},
\newblock \bibinfo{journal}{\apj} \bibinfo{volume}{621} (\bibinfo{year}{2005})
  \bibinfo{pages}{643--650}. \DOIprefix\doi{10.1086/427679}.
  \href{http://arxiv.org/abs/astro-ph/0406527}{{\tt arXiv:astro-ph/0406527}}.
\bibitem[{{Beygu} et~al.(2016){Beygu}, {Kreckel}, {van der Hulst}, {Jarrett},
  {Peletier}, {van de Weygaert}, {van Gorkom}, and {Aragon-Calvo}}]{beygu16}
\bibinfo{author}{B.~{Beygu}}, \bibinfo{author}{K.~{Kreckel}},
  \bibinfo{author}{J.~M. {van der Hulst}}, \bibinfo{author}{T.~H. {Jarrett}},
  \bibinfo{author}{R.~{Peletier}}, \bibinfo{author}{R.~{van de Weygaert}},
  \bibinfo{author}{J.~H. {van Gorkom}}, \bibinfo{author}{M.~A. {Aragon-Calvo}},
\newblock \bibinfo{title}{{The void galaxy survey: Star formation properties}},
\newblock \bibinfo{journal}{\mnras} \bibinfo{volume}{458}
  (\bibinfo{year}{2016}) \bibinfo{pages}{394--409}.
  \DOIprefix\doi{10.1093/mnras/stw280}.
  \href{http://arxiv.org/abs/1601.08228}{{\tt arXiv:1601.08228}}.
\bibitem[{{von Benda-Beckmann} and {M{\"u}ller}(2008)}]{vonbenda08}
\bibinfo{author}{A.~M. {von Benda-Beckmann}},
  \bibinfo{author}{V.~{M{\"u}ller}},
\newblock \bibinfo{title}{{Void statistics and void galaxies in the 2dF Galaxy
  Redshift Survey}},
\newblock \bibinfo{journal}{\mnras} \bibinfo{volume}{384}
  (\bibinfo{year}{2008}) \bibinfo{pages}{1189--1199}.
  \DOIprefix\doi{10.1111/j.1365-2966.2007.12789.x}.
\bibitem[{{Hoyle} et~al.(2012){Hoyle}, {Vogeley}, and {Pan}}]{hoyle12}
\bibinfo{author}{F.~{Hoyle}}, \bibinfo{author}{M.~S. {Vogeley}},
  \bibinfo{author}{D.~{Pan}},
\newblock \bibinfo{title}{{Photometric properties of void galaxies in the Sloan
  Digital Sky Survey Data Release 7}},
\newblock \bibinfo{journal}{\mnras} \bibinfo{volume}{426}
  (\bibinfo{year}{2012}) \bibinfo{pages}{3041--3050}.
  \DOIprefix\doi{10.1111/j.1365-2966.2012.21943.x}.
  \href{http://arxiv.org/abs/1205.1843}{{\tt arXiv:1205.1843}}.
\bibitem[{{Rojas} et~al.(2004){Rojas}, {Vogeley}, {Hoyle}, and
  {Brinkmann}}]{rojas04}
\bibinfo{author}{R.~R. {Rojas}}, \bibinfo{author}{M.~S. {Vogeley}},
  \bibinfo{author}{F.~{Hoyle}}, \bibinfo{author}{J.~{Brinkmann}},
\newblock \bibinfo{title}{{Photometric Properties of Void Galaxies in the Sloan
  Digital Sky Survey}},
\newblock \bibinfo{journal}{\apj} \bibinfo{volume}{617} (\bibinfo{year}{2004})
  \bibinfo{pages}{50--63}. \DOIprefix\doi{10.1086/425225}.
  \href{http://arxiv.org/abs/astro-ph/0307274}{{\tt arXiv:astro-ph/0307274}}.
\bibitem[{{Rojas} et~al.(2005){Rojas}, {Vogeley}, {Hoyle}, and
  {Brinkmann}}]{rojas05}
\bibinfo{author}{R.~R. {Rojas}}, \bibinfo{author}{M.~S. {Vogeley}},
  \bibinfo{author}{F.~{Hoyle}}, \bibinfo{author}{J.~{Brinkmann}},
\newblock \bibinfo{title}{{Spectroscopic Properties of Void Galaxies in the
  Sloan Digital Sky Survey}},
\newblock \bibinfo{journal}{\apj} \bibinfo{volume}{624} (\bibinfo{year}{2005})
  \bibinfo{pages}{571--585}. \DOIprefix\doi{10.1086/428476}.
  \href{http://arxiv.org/abs/astro-ph/0409074}{{\tt arXiv:astro-ph/0409074}}.
\bibitem[{{Croton} and {Farrar}(2008)}]{croton08}
\bibinfo{author}{D.~J. {Croton}}, \bibinfo{author}{G.~R. {Farrar}},
\newblock \bibinfo{title}{{Where do `red and dead' early-type void galaxies
  come from?}},
\newblock \bibinfo{journal}{\mnras} \bibinfo{volume}{386}
  (\bibinfo{year}{2008}) \bibinfo{pages}{2285--2289}.
  \DOIprefix\doi{10.1111/j.1365-2966.2008.13204.x}.
  \href{http://arxiv.org/abs/0801.2771}{{\tt arXiv:0801.2771}}.
\bibitem[{{Betancort-Rijo} et~al.(2009){Betancort-Rijo}, {Patiri}, {Prada}, and
  {Romano}}]{rijo09}
\bibinfo{author}{J.~{Betancort-Rijo}}, \bibinfo{author}{S.~G. {Patiri}},
  \bibinfo{author}{F.~{Prada}}, \bibinfo{author}{A.~E. {Romano}},
\newblock \bibinfo{title}{{The statistics of voids as a tool to constrain
  cosmological parameters: {\ensuremath{\sigma}}$_{8}$ and
  {\ensuremath{\Gamma}}}},
\newblock \bibinfo{journal}{\mnras} \bibinfo{volume}{400}
  (\bibinfo{year}{2009}) \bibinfo{pages}{1835--1849}.
  \DOIprefix\doi{10.1111/j.1365-2966.2009.15567.x}.
  \href{http://arxiv.org/abs/0901.1609}{{\tt arXiv:0901.1609}}.
\bibitem[{{Croton} et~al.(2005){Croton}, {Farrar}, {Norberg}, {Colless},
  {Peacock}, {Baldry}, {Baugh}, {Bland-Hawthorn}, {Bridges}, {Cannon}, {Cole},
  {Collins}, {Couch}, {Dalton}, {De Propris}, {Driver}, {Efstathiou}, {Ellis},
  {Frenk}, {Glazebrook}, {Jackson}, {Lahav}, {Lewis}, {Lumsden}, {Maddox},
  {Madgwick}, {Peterson}, {Sutherland}, and {Taylor}}]{croton05}
\bibinfo{author}{D.~J. {Croton}}, \bibinfo{author}{G.~R. {Farrar}},
  \bibinfo{author}{P.~{Norberg}}, \bibinfo{author}{M.~{Colless}},
  \bibinfo{author}{J.~A. {Peacock}}, \bibinfo{author}{I.~K. {Baldry}},
  \bibinfo{author}{C.~M. {Baugh}}, \bibinfo{author}{J.~{Bland-Hawthorn}},
  \bibinfo{author}{T.~{Bridges}}, \bibinfo{author}{R.~{Cannon}},
  \bibinfo{author}{S.~{Cole}}, \bibinfo{author}{C.~{Collins}},
  \bibinfo{author}{W.~{Couch}}, \bibinfo{author}{G.~{Dalton}},
  \bibinfo{author}{R.~{De Propris}}, \bibinfo{author}{S.~P. {Driver}},
  \bibinfo{author}{G.~{Efstathiou}}, \bibinfo{author}{R.~S. {Ellis}},
  \bibinfo{author}{C.~S. {Frenk}}, \bibinfo{author}{K.~{Glazebrook}},
  \bibinfo{author}{C.~{Jackson}}, \bibinfo{author}{O.~{Lahav}},
  \bibinfo{author}{I.~{Lewis}}, \bibinfo{author}{S.~{Lumsden}},
  \bibinfo{author}{S.~{Maddox}}, \bibinfo{author}{D.~{Madgwick}},
  \bibinfo{author}{B.~A. {Peterson}}, \bibinfo{author}{W.~{Sutherland}},
  \bibinfo{author}{K.~{Taylor}},
\newblock \bibinfo{title}{{The 2dF Galaxy Redshift Survey: luminosity functions
  by density environment and galaxy type}},
\newblock \bibinfo{journal}{\mnras} \bibinfo{volume}{356}
  (\bibinfo{year}{2005}) \bibinfo{pages}{1155--1167}.
  \DOIprefix\doi{10.1111/j.1365-2966.2004.08546.x}.
  \href{http://arxiv.org/abs/astro-ph/0407537}{{\tt arXiv:astro-ph/0407537}}.
\bibitem[{{Moorman} et~al.(2014){Moorman}, {Vogeley}, {Hoyle}, {Pan}, {Haynes},
  and {Giovanelli}}]{moor14}
\bibinfo{author}{C.~M. {Moorman}}, \bibinfo{author}{M.~S. {Vogeley}},
  \bibinfo{author}{F.~{Hoyle}}, \bibinfo{author}{D.~C. {Pan}},
  \bibinfo{author}{M.~P. {Haynes}}, \bibinfo{author}{R.~{Giovanelli}},
\newblock \bibinfo{title}{{The H I mass function and velocity width function of
  void galaxies in the Arecibo Legacy Fast ALFA Survey}},
\newblock \bibinfo{journal}{\mnras} \bibinfo{volume}{444}
  (\bibinfo{year}{2014}) \bibinfo{pages}{3559--3570}.
  \DOIprefix\doi{10.1093/mnras/stu1674}.
  \href{http://arxiv.org/abs/1408.3392}{{\tt arXiv:1408.3392}}.
\bibitem[{{Moorman} et~al.(2015){Moorman}, {Vogeley}, {Hoyle}, {Pan}, {Haynes},
  and {Giovanelli}}]{moor15}
\bibinfo{author}{C.~M. {Moorman}}, \bibinfo{author}{M.~S. {Vogeley}},
  \bibinfo{author}{F.~{Hoyle}}, \bibinfo{author}{D.~C. {Pan}},
  \bibinfo{author}{M.~P. {Haynes}}, \bibinfo{author}{R.~{Giovanelli}},
\newblock \bibinfo{title}{{The Optical Luminosity Function of Void Galaxies in
  the SDSS and ALFALFA Surveys}},
\newblock \bibinfo{journal}{\apj} \bibinfo{volume}{810} (\bibinfo{year}{2015})
  \bibinfo{pages}{108}. \DOIprefix\doi{10.1088/0004-637X/810/2/108}.
  \href{http://arxiv.org/abs/1508.04199}{{\tt arXiv:1508.04199}}.
\bibitem[{{Moorman} et~al.(2016){Moorman}, {Moreno}, {White}, {Vogeley},
  {Hoyle}, {Giovanelli}, and {Haynes}}]{moor16}
\bibinfo{author}{C.~M. {Moorman}}, \bibinfo{author}{J.~{Moreno}},
  \bibinfo{author}{A.~{White}}, \bibinfo{author}{M.~S. {Vogeley}},
  \bibinfo{author}{F.~{Hoyle}}, \bibinfo{author}{R.~{Giovanelli}},
  \bibinfo{author}{M.~P. {Haynes}},
\newblock \bibinfo{title}{{On the Star Formation Properties of Void Galaxies}},
\newblock \bibinfo{journal}{\apj} \bibinfo{volume}{831} (\bibinfo{year}{2016})
  \bibinfo{pages}{118}. \DOIprefix\doi{10.3847/0004-637X/831/2/118}.
  \href{http://arxiv.org/abs/1601.04092}{{\tt arXiv:1601.04092}}.
\bibitem[{{Furlanetto} and {Piran}(2006)}]{furl06}
\bibinfo{author}{S.~R. {Furlanetto}}, \bibinfo{author}{T.~{Piran}},
\newblock \bibinfo{title}{{The evidence of absence: galaxy voids in the
  excursion set formalism}},
\newblock \bibinfo{journal}{\mnras} \bibinfo{volume}{366}
  (\bibinfo{year}{2006}) \bibinfo{pages}{467--479}.
  \DOIprefix\doi{10.1111/j.1365-2966.2005.09862.x}.
  \href{http://arxiv.org/abs/astro-ph/0509148}{{\tt arXiv:astro-ph/0509148}}.
\bibitem[{{Sheth} et~al.(2001){Sheth}, {Mo}, and {Tormen}}]{sheth01}
\bibinfo{author}{R.~K. {Sheth}}, \bibinfo{author}{H.~J. {Mo}},
  \bibinfo{author}{G.~{Tormen}},
\newblock \bibinfo{title}{{Ellipsoidal collapse and an improved model for the
  number and spatial distribution of dark matter haloes}},
\newblock \bibinfo{journal}{\mnras} \bibinfo{volume}{323}
  (\bibinfo{year}{2001}) \bibinfo{pages}{1--12}.
  \DOIprefix\doi{10.1046/j.1365-8711.2001.04006.x}.
  \href{http://arxiv.org/abs/astro-ph/9907024}{{\tt arXiv:astro-ph/9907024}}.
\bibitem[{{Samui}(2014)}]{samui14}
\bibinfo{author}{S.~{Samui}},
\newblock \bibinfo{title}{{Star formation in high redshift galaxies including
  supernova feedback: Effect on stellar mass and luminosity functions}},
\newblock \bibinfo{journal}{\na} \bibinfo{volume}{30} (\bibinfo{year}{2014})
  \bibinfo{pages}{89--99}. \DOIprefix\doi{10.1016/j.newast.2014.01.010}.
  \href{http://arxiv.org/abs/1401.7675}{{\tt arXiv:1401.7675}}.
\bibitem[{{Neronov} and {Vovk}(2010)}]{nero10}
\bibinfo{author}{A.~{Neronov}}, \bibinfo{author}{I.~{Vovk}},
\newblock \bibinfo{title}{{Evidence for Strong Extragalactic Magnetic Fields
  from Fermi Observations of TeV Blazars}},
\newblock \bibinfo{journal}{Science} \bibinfo{volume}{328}
  (\bibinfo{year}{2010}) \bibinfo{pages}{73}.
  \DOIprefix\doi{10.1126/science.1184192}.
  \href{http://arxiv.org/abs/1006.3504}{{\tt arXiv:1006.3504}}.
\bibitem[{{Beck}(2001)}]{rbeck01}
\bibinfo{author}{R.~{Beck}},
\newblock \bibinfo{title}{{Galactic and Extragalactic Magnetic Fields}},
\newblock \bibinfo{journal}{\ssr} \bibinfo{volume}{99} (\bibinfo{year}{2001})
  \bibinfo{pages}{243--260}. \href{http://arxiv.org/abs/astro-ph/0012402}{{\tt
  arXiv:astro-ph/0012402}}.
\bibitem[{{Beck} and {Wielebinski}(2013)}]{beckwie13}
\bibinfo{author}{R.~{Beck}}, \bibinfo{author}{R.~{Wielebinski}},
  \bibinfo{title}{{Magnetic Fields in Galaxies}}, \bibinfo{year}{2013}, p.
  \bibinfo{pages}{641}. \DOIprefix\doi{10.1007/978-94-007-5612-0_13}.
\bibitem[{{Subramanian}(2016)}]{subr16}
\bibinfo{author}{K.~{Subramanian}},
\newblock \bibinfo{title}{{The origin, evolution and signatures of primordial
  magnetic fields}},
\newblock \bibinfo{journal}{Reports on Progress in Physics}
  \bibinfo{volume}{79} (\bibinfo{year}{2016}) \bibinfo{pages}{076901}.
  \DOIprefix\doi{10.1088/0034-4885/79/7/076901}.
  \href{http://arxiv.org/abs/1504.02311}{{\tt arXiv:1504.02311}}.
\bibitem[{{Durrer} and {Neronov}(2013)}]{durrer13}
\bibinfo{author}{R.~{Durrer}}, \bibinfo{author}{A.~{Neronov}},
\newblock \bibinfo{title}{{Cosmological magnetic fields: their generation,
  evolution and observation}},
\newblock \bibinfo{journal}{\aapr} \bibinfo{volume}{21} (\bibinfo{year}{2013})
  \bibinfo{pages}{62}. \DOIprefix\doi{10.1007/s00159-013-0062-7}.
  \href{http://arxiv.org/abs/1303.7121}{{\tt arXiv:1303.7121}}.
\bibitem[{{Turner} and {Widrow}(1988)}]{turner88}
\bibinfo{author}{M.~S. {Turner}}, \bibinfo{author}{L.~M. {Widrow}},
\newblock \bibinfo{title}{{Inflation-produced, large-scale magnetic fields}},
\newblock \bibinfo{journal}{\prd} \bibinfo{volume}{37} (\bibinfo{year}{1988})
  \bibinfo{pages}{2743--2754}. \DOIprefix\doi{10.1103/PhysRevD.37.2743}.
\bibitem[{{Banerjee} and {Jedamzik}(2004)}]{banerjee04}
\bibinfo{author}{R.~{Banerjee}}, \bibinfo{author}{K.~{Jedamzik}},
\newblock \bibinfo{title}{{Evolution of cosmic magnetic fields: From the very
  early Universe, to recombination, to the present}},
\newblock \bibinfo{journal}{\prd} \bibinfo{volume}{70} (\bibinfo{year}{2004})
  \bibinfo{pages}{123003}. \DOIprefix\doi{10.1103/PhysRevD.70.123003}.
  \href{http://arxiv.org/abs/astro-ph/0410032}{{\tt arXiv:astro-ph/0410032}}.
\bibitem[{{Ichiki} et~al.(2006){Ichiki}, {Takahashi}, {Ohno}, {Hanayama}, and
  {Sugiyama}}]{ichiki06}
\bibinfo{author}{K.~{Ichiki}}, \bibinfo{author}{K.~{Takahashi}},
  \bibinfo{author}{H.~{Ohno}}, \bibinfo{author}{H.~{Hanayama}},
  \bibinfo{author}{N.~{Sugiyama}},
\newblock \bibinfo{title}{{Cosmological Magnetic Field: A Fossil of Density
  Perturbations in the Early Universe}},
\newblock \bibinfo{journal}{Science} \bibinfo{volume}{311}
  (\bibinfo{year}{2006}) \bibinfo{pages}{827--829}.
  \DOIprefix\doi{10.1126/science.1120690}.
  \href{http://arxiv.org/abs/astro-ph/0603631}{{\tt arXiv:astro-ph/0603631}}.
\bibitem[{{Gnedin} et~al.(2000){Gnedin}, {Ferrara}, and {Zweibel}}]{gnedin00}
\bibinfo{author}{N.~Y. {Gnedin}}, \bibinfo{author}{A.~{Ferrara}},
  \bibinfo{author}{E.~G. {Zweibel}},
\newblock \bibinfo{title}{{Generation of the Primordial Magnetic Fields during
  Cosmological Reionization}},
\newblock \bibinfo{journal}{\apj} \bibinfo{volume}{539} (\bibinfo{year}{2000})
  \bibinfo{pages}{505--516}. \DOIprefix\doi{10.1086/309272}.
  \href{http://arxiv.org/abs/astro-ph/0001066}{{\tt arXiv:astro-ph/0001066}}.
\bibitem[{{Medvedev}(2007)}]{medve07}
\bibinfo{author}{M.~V. {Medvedev}},
\newblock \bibinfo{title}{{Weibel Turbulence in Laboratory Experiments and
  GRB/SN Shocks}},
\newblock \bibinfo{journal}{\apss} \bibinfo{volume}{307} (\bibinfo{year}{2007})
  \bibinfo{pages}{245--250}. \DOIprefix\doi{10.1007/s10509-006-9288-4}.
\bibitem[{{Martin-Alvarez} et~al.(2017){Martin-Alvarez}, {Planelles}, and
  {Quilis}}]{martin17}
\bibinfo{author}{S.~{Martin-Alvarez}}, \bibinfo{author}{S.~{Planelles}},
  \bibinfo{author}{V.~{Quilis}},
\newblock \bibinfo{title}{{On the interplay between cosmological shock waves
  and their environment}},
\newblock \bibinfo{journal}{\apss} \bibinfo{volume}{362} (\bibinfo{year}{2017})
  \bibinfo{pages}{91}. \DOIprefix\doi{10.1007/s10509-017-3066-3}.
\bibitem[{{Miniati} and {Bell}(2011)}]{miniati11}
\bibinfo{author}{F.~{Miniati}}, \bibinfo{author}{A.~R. {Bell}},
\newblock \bibinfo{title}{{Resistive Magnetic Field Generation at Cosmic
  Dawn}},
\newblock \bibinfo{journal}{\apj} \bibinfo{volume}{729} (\bibinfo{year}{2011})
  \bibinfo{pages}{73}. \DOIprefix\doi{10.1088/0004-637X/729/1/73}.
  \href{http://arxiv.org/abs/1001.2011}{{\tt arXiv:1001.2011}}.
\bibitem[{{Madau} et~al.(2001){Madau}, {Ferrara}, and {Rees}}]{madau01}
\bibinfo{author}{P.~{Madau}}, \bibinfo{author}{A.~{Ferrara}},
  \bibinfo{author}{M.~J. {Rees}},
\newblock \bibinfo{title}{{Early Metal Enrichment of the Intergalactic Medium
  by Pregalactic Outflows}},
\newblock \bibinfo{journal}{\apj} \bibinfo{volume}{555} (\bibinfo{year}{2001})
  \bibinfo{pages}{92--105}. \DOIprefix\doi{10.1086/321474}.
  \href{http://arxiv.org/abs/astro-ph/0010158}{{\tt arXiv:astro-ph/0010158}}.
\bibitem[{{Furlanetto} et~al.(2003){Furlanetto}, {Schaye}, {Springel}, and
  {Hernquist}}]{furl03}
\bibinfo{author}{S.~R. {Furlanetto}}, \bibinfo{author}{J.~{Schaye}},
  \bibinfo{author}{V.~{Springel}}, \bibinfo{author}{L.~{Hernquist}},
\newblock \bibinfo{title}{{Mapping the Cosmic Web with Ly{$\alpha$} Emission}},
\newblock \bibinfo{journal}{\apjl} \bibinfo{volume}{599} (\bibinfo{year}{2003})
  \bibinfo{pages}{L1--L4}. \DOIprefix\doi{10.1086/381140}.
  \href{http://arxiv.org/abs/astro-ph/0311006}{{\tt arXiv:astro-ph/0311006}}.
\bibitem[{{Scannapieco} et~al.(2005){Scannapieco}, {Tissera}, {White}, and
  {Springel}}]{scan05}
\bibinfo{author}{C.~{Scannapieco}}, \bibinfo{author}{P.~B. {Tissera}},
  \bibinfo{author}{S.~D.~M. {White}}, \bibinfo{author}{V.~{Springel}},
\newblock \bibinfo{title}{{Feedback and metal enrichment in cosmological
  smoothed particle hydrodynamics simulations - I. A model for chemical
  enrichment}},
\newblock \bibinfo{journal}{\mnras} \bibinfo{volume}{364}
  (\bibinfo{year}{2005}) \bibinfo{pages}{552--564}.
  \DOIprefix\doi{10.1111/j.1365-2966.2005.09574.x}.
  \href{http://arxiv.org/abs/astro-ph/0505440}{{\tt arXiv:astro-ph/0505440}}.
\bibitem[{{Bertone} et~al.(2006){Bertone}, {Vogt}, and {En{\ss}lin}}]{bert06}
\bibinfo{author}{S.~{Bertone}}, \bibinfo{author}{C.~{Vogt}},
  \bibinfo{author}{T.~{En{\ss}lin}},
\newblock \bibinfo{title}{{Magnetic field seeding by galactic winds}},
\newblock \bibinfo{journal}{\mnras} \bibinfo{volume}{370}
  (\bibinfo{year}{2006}) \bibinfo{pages}{319--330}.
  \DOIprefix\doi{10.1111/j.1365-2966.2006.10474.x}.
  \href{http://arxiv.org/abs/astro-ph/0604462}{{\tt arXiv:astro-ph/0604462}}.
\bibitem[{{Samui} et~al.(2008){Samui}, {Subramanian}, and {Srianand
  }}]{samui08}
\bibinfo{author}{S.~{Samui}}, \bibinfo{author}{K.~{Subramanian}},
  \bibinfo{author}{R.~{Srianand }},
\newblock \bibinfo{title}{{Constrained semi-analytical models of galactic
  outflows}},
\newblock \bibinfo{journal}{\mnras} \bibinfo{volume}{385}
  (\bibinfo{year}{2008}) \bibinfo{pages}{783--808}.
  \DOIprefix\doi{10.1111/j.1365-2966.2008.12932.x}.
  \href{http://arxiv.org/abs/0801.1401}{{\tt arXiv:0801.1401}}.
\bibitem[{{Beck} et~al.(2013){Beck}, {Hanasz}, {Lesch}, {Remus}, and
  {Stasyszyn}}]{beck13}
\bibinfo{author}{A.~M. {Beck}}, \bibinfo{author}{M.~{Hanasz}},
  \bibinfo{author}{H.~{Lesch}}, \bibinfo{author}{R.-S. {Remus}},
  \bibinfo{author}{F.~A. {Stasyszyn}},
\newblock \bibinfo{title}{{On the magnetic fields in voids}},
\newblock \bibinfo{journal}{\mnras} \bibinfo{volume}{429}
  (\bibinfo{year}{2013}) \bibinfo{pages}{L60--L64}.
  \DOIprefix\doi{10.1093/mnrasl/sls026}.
  \href{http://arxiv.org/abs/1210.8360}{{\tt arXiv:1210.8360}}.
\bibitem[{{Samui} et~al.(2018){Samui}, {Subramanian}, and {Srianand}}]{samui18}
\bibinfo{author}{S.~{Samui}}, \bibinfo{author}{K.~{Subramanian}},
  \bibinfo{author}{R.~{Srianand}},
\newblock \bibinfo{title}{{Efficient cold outflows driven by cosmic rays in
  high-redshift galaxies and their global effects on the IGM}},
\newblock \bibinfo{journal}{\mnras} \bibinfo{volume}{476}
  (\bibinfo{year}{2018}) \bibinfo{pages}{1680--1695}.
  \DOIprefix\doi{10.1093/mnras/sty287}.
  \href{http://arxiv.org/abs/1706.01890}{{\tt arXiv:1706.01890}}.
\bibitem[{{Planck Collaboration} et~al.(2016){Planck Collaboration}, {Ade},
  {Aghanim}, {Arnaud}, {Ashdown}, {Aumont}, {Baccigalupi}, {Banday},
  {Barreiro}, {Bartlett}, {Bartolo}, {Battaner}, {Battye}, {Benabed},
  {Beno{\^\i}t}, {Benoit-L{\'e}vy}, {Bernard}, {Bersanelli}, {Bielewicz},
  {Bock}, {Bonaldi}, {Bonavera}, {Bond}, {Borrill}, {Bouchet}, {Boulanger},
  {Bucher}, {Burigana}, {Butler}, {Calabrese}, {Cardoso}, {Catalano},
  {Challinor}, {Chamballu}, {Chary}, {Chiang}, {Chluba}, {Christensen},
  {Church}, {Clements}, {Colombi}, {Colombo}, {Combet}, {Coulais}, {Crill},
  {Curto}, {Cuttaia}, {Danese}, {Davies}, {Davis}, {de Bernardis}, {de Rosa},
  {de Zotti}, {Delabrouille}, {D{\'e}sert}, {Di Valentino}, {Dickinson},
  {Diego}, {Dolag}, {Dole}, {Donzelli}, {Dor{\'e}}, {Douspis}, {Ducout},
  {Dunkley}, {Dupac}, {Efstathiou}, {Elsner}, {En{\ss}lin}, {Eriksen},
  {Farhang}, {Fergusson}, {Finelli}, {Forni}, {Frailis}, {Fraisse},
  {Franceschi}, {Frejsel}, {Galeotta}, {Galli}, {Ganga}, {Gauthier}, {Gerbino},
  {Ghosh}, {Giard}, {Giraud-H{\'e}raud}, {Giusarma}, {Gjerl{\o}w},
  {Gonz{\'a}lez-Nuevo}, {G{\'o}rski}, {Gratton}, {Gregorio}, {Gruppuso},
  {Gudmundsson}, {Hamann}, {Hansen}, {Hanson}, {Harrison}, {Helou},
  {Henrot-Versill{\'e}}, {Hern{\'a}ndez-Monteagudo}, {Herranz}, {Hildebrand t},
  {Hivon}, {Hobson}, {Holmes}, {Hornstrup}, {Hovest}, {Huang}, {Huffenberger},
  {Hurier}, {Jaffe}, {Jaffe}, {Jones}, {Juvela}, {Keih{\"a}nen}, {Keskitalo},
  {Kisner}, {Kneissl}, {Knoche}, {Knox}, {Kunz}, {Kurki-Suonio}, {Lagache},
  {L{\"a}hteenm{\"a}ki}, {Lamarre}, {Lasenby}, {Lattanzi}, {Lawrence}, {Leahy},
  {Leonardi}, {Lesgourgues}, {Levrier}, {Lewis}, {Liguori}, {Lilje},
  {Linden-V{\o}rnle}, {L{\'o}pez-Caniego}, {Lubin}, {Mac{\'\i}as-P{\'e}rez},
  {Maggio}, {Maino}, {Mandolesi}, {Mangilli}, {Marchini}, {Maris}, {Martin},
  {Martinelli}, {Mart{\'\i}nez-Gonz{\'a}lez}, {Masi}, {Matarrese}, {McGehee},
  {Meinhold}, {Melchiorri}, {Melin}, {Mendes}, {Mennella}, {Migliaccio},
  {Millea}, {Mitra}, {Miville-Desch{\^e}nes}, {Moneti}, {Montier}, {Morgante},
  {Mortlock}, {Moss}, {Munshi}, {Murphy}, {Naselsky}, {Nati}, {Natoli},
  {Netterfield}, {N{\o}rgaard-Nielsen}, {Noviello}, {Novikov}, {Novikov},
  {Oxborrow}, {Paci}, {Pagano}, {Pajot}, {Paladini}, {Paoletti}, {Partridge},
  {Pasian}, {Patanchon}, {Pearson}, {Perdereau}, {Perotto}, {Perrotta},
  {Pettorino}, {Piacentini}, {Piat}, {Pierpaoli}, {Pietrobon}, {Plaszczynski},
  {Pointecouteau}, {Polenta}, {Popa}, {Pratt}, {Pr{\'e}zeau}, {Prunet},
  {Puget}, {Rachen}, {Reach}, {Rebolo}, {Reinecke}, {Remazeilles}, {Renault},
  {Renzi}, {Ristorcelli}, {Rocha}, {Rosset}, {Rossetti}, {Roudier},
  {Rouill{\'e} d'Orfeuil}, {Rowan-Robinson}, {Rubi{\~n}o-Mart{\'\i}n},
  {Rusholme}, {Said}, {Salvatelli}, {Salvati}, {Sandri}, {Santos},
  {Savelainen}, {Savini}, {Scott}, {Seiffert}, {Serra}, {Shellard}, {Spencer},
  {Spinelli}, {Stolyarov}, {Stompor}, {Sudiwala}, {Sunyaev}, {Sutton},
  {Suur-Uski}, {Sygnet}, {Tauber}, {Terenzi}, {Toffolatti}, {Tomasi},
  {Tristram}, {Trombetti}, {Tucci}, {Tuovinen}, {T{\"u}rler}, {Umana},
  {Valenziano}, {Valiviita}, {Van Tent}, {Vielva}, {Villa}, {Wade}, {Wandelt},
  {Wehus}, {White}, {White}, {Wilkinson}, {Yvon}, {Zacchei}, and
  {Zonca}}]{2016A&A...594A..13P}
\bibinfo{author}{{Planck Collaboration}}, \bibinfo{author}{P.~A.~R. {Ade}},
  \bibinfo{author}{N.~{Aghanim}}, \bibinfo{author}{M.~{Arnaud}},
  \bibinfo{author}{M.~{Ashdown}}, \bibinfo{author}{J.~{Aumont}},
  \bibinfo{author}{C.~{Baccigalupi}}, \bibinfo{author}{A.~J. {Banday}},
  \bibinfo{author}{R.~B. {Barreiro}}, \bibinfo{author}{J.~G. {Bartlett}},
  \bibinfo{author}{N.~{Bartolo}}, \bibinfo{author}{E.~{Battaner}},
  \bibinfo{author}{R.~{Battye}}, \bibinfo{author}{K.~{Benabed}},
  \bibinfo{author}{A.~{Beno{\^\i}t}}, \bibinfo{author}{A.~{Benoit-L{\'e}vy}},
  \bibinfo{author}{J.~P. {Bernard}}, \bibinfo{author}{M.~{Bersanelli}},
  \bibinfo{author}{P.~{Bielewicz}}, \bibinfo{author}{J.~J. {Bock}},
  \bibinfo{author}{A.~{Bonaldi}}, \bibinfo{author}{L.~{Bonavera}},
  \bibinfo{author}{J.~R. {Bond}}, \bibinfo{author}{J.~{Borrill}},
  \bibinfo{author}{F.~R. {Bouchet}}, \bibinfo{author}{F.~{Boulanger}},
  \bibinfo{author}{M.~{Bucher}}, \bibinfo{author}{C.~{Burigana}},
  \bibinfo{author}{R.~C. {Butler}}, \bibinfo{author}{E.~{Calabrese}},
  \bibinfo{author}{J.~F. {Cardoso}}, \bibinfo{author}{A.~{Catalano}},
  \bibinfo{author}{A.~{Challinor}}, \bibinfo{author}{A.~{Chamballu}},
  \bibinfo{author}{R.~R. {Chary}}, \bibinfo{author}{H.~C. {Chiang}},
  \bibinfo{author}{J.~{Chluba}}, \bibinfo{author}{P.~R. {Christensen}},
  \bibinfo{author}{S.~{Church}}, \bibinfo{author}{D.~L. {Clements}},
  \bibinfo{author}{S.~{Colombi}}, \bibinfo{author}{L.~P.~L. {Colombo}},
  \bibinfo{author}{C.~{Combet}}, \bibinfo{author}{A.~{Coulais}},
  \bibinfo{author}{B.~P. {Crill}}, \bibinfo{author}{A.~{Curto}},
  \bibinfo{author}{F.~{Cuttaia}}, \bibinfo{author}{L.~{Danese}},
  \bibinfo{author}{R.~D. {Davies}}, \bibinfo{author}{R.~J. {Davis}},
  \bibinfo{author}{P.~{de Bernardis}}, \bibinfo{author}{A.~{de Rosa}},
  \bibinfo{author}{G.~{de Zotti}}, \bibinfo{author}{J.~{Delabrouille}},
  \bibinfo{author}{F.~X. {D{\'e}sert}}, \bibinfo{author}{E.~{Di Valentino}},
  \bibinfo{author}{C.~{Dickinson}}, \bibinfo{author}{J.~M. {Diego}},
  \bibinfo{author}{K.~{Dolag}}, \bibinfo{author}{H.~{Dole}},
  \bibinfo{author}{S.~{Donzelli}}, \bibinfo{author}{O.~{Dor{\'e}}},
  \bibinfo{author}{M.~{Douspis}}, \bibinfo{author}{A.~{Ducout}},
  \bibinfo{author}{J.~{Dunkley}}, \bibinfo{author}{X.~{Dupac}},
  \bibinfo{author}{G.~{Efstathiou}}, \bibinfo{author}{F.~{Elsner}},
  \bibinfo{author}{T.~A. {En{\ss}lin}}, \bibinfo{author}{H.~K. {Eriksen}},
  \bibinfo{author}{M.~{Farhang}}, \bibinfo{author}{J.~{Fergusson}},
  \bibinfo{author}{F.~{Finelli}}, \bibinfo{author}{O.~{Forni}},
  \bibinfo{author}{M.~{Frailis}}, \bibinfo{author}{A.~A. {Fraisse}},
  \bibinfo{author}{E.~{Franceschi}}, \bibinfo{author}{A.~{Frejsel}},
  \bibinfo{author}{S.~{Galeotta}}, \bibinfo{author}{S.~{Galli}},
  \bibinfo{author}{K.~{Ganga}}, \bibinfo{author}{C.~{Gauthier}},
  \bibinfo{author}{M.~{Gerbino}}, \bibinfo{author}{T.~{Ghosh}},
  \bibinfo{author}{M.~{Giard}}, \bibinfo{author}{Y.~{Giraud-H{\'e}raud}},
  \bibinfo{author}{E.~{Giusarma}}, \bibinfo{author}{E.~{Gjerl{\o}w}},
  \bibinfo{author}{J.~{Gonz{\'a}lez-Nuevo}}, \bibinfo{author}{K.~M.
  {G{\'o}rski}}, \bibinfo{author}{S.~{Gratton}},
  \bibinfo{author}{A.~{Gregorio}}, \bibinfo{author}{A.~{Gruppuso}},
  \bibinfo{author}{J.~E. {Gudmundsson}}, \bibinfo{author}{J.~{Hamann}},
  \bibinfo{author}{F.~K. {Hansen}}, \bibinfo{author}{D.~{Hanson}},
  \bibinfo{author}{D.~L. {Harrison}}, \bibinfo{author}{G.~{Helou}},
  \bibinfo{author}{S.~{Henrot-Versill{\'e}}},
  \bibinfo{author}{C.~{Hern{\'a}ndez-Monteagudo}},
  \bibinfo{author}{D.~{Herranz}}, \bibinfo{author}{S.~R. {Hildebrand t}},
  \bibinfo{author}{E.~{Hivon}}, \bibinfo{author}{M.~{Hobson}},
  \bibinfo{author}{W.~A. {Holmes}}, \bibinfo{author}{A.~{Hornstrup}},
  \bibinfo{author}{W.~{Hovest}}, \bibinfo{author}{Z.~{Huang}},
  \bibinfo{author}{K.~M. {Huffenberger}}, \bibinfo{author}{G.~{Hurier}},
  \bibinfo{author}{A.~H. {Jaffe}}, \bibinfo{author}{T.~R. {Jaffe}},
  \bibinfo{author}{W.~C. {Jones}}, \bibinfo{author}{M.~{Juvela}},
  \bibinfo{author}{E.~{Keih{\"a}nen}}, \bibinfo{author}{R.~{Keskitalo}},
  \bibinfo{author}{T.~S. {Kisner}}, \bibinfo{author}{R.~{Kneissl}},
  \bibinfo{author}{J.~{Knoche}}, \bibinfo{author}{L.~{Knox}},
  \bibinfo{author}{M.~{Kunz}}, \bibinfo{author}{H.~{Kurki-Suonio}},
  \bibinfo{author}{G.~{Lagache}}, \bibinfo{author}{A.~{L{\"a}hteenm{\"a}ki}},
  \bibinfo{author}{J.~M. {Lamarre}}, \bibinfo{author}{A.~{Lasenby}},
  \bibinfo{author}{M.~{Lattanzi}}, \bibinfo{author}{C.~R. {Lawrence}},
  \bibinfo{author}{J.~P. {Leahy}}, \bibinfo{author}{R.~{Leonardi}},
  \bibinfo{author}{J.~{Lesgourgues}}, \bibinfo{author}{F.~{Levrier}},
  \bibinfo{author}{A.~{Lewis}}, \bibinfo{author}{M.~{Liguori}},
  \bibinfo{author}{P.~B. {Lilje}}, \bibinfo{author}{M.~{Linden-V{\o}rnle}},
  \bibinfo{author}{M.~{L{\'o}pez-Caniego}}, \bibinfo{author}{P.~M. {Lubin}},
  \bibinfo{author}{J.~F. {Mac{\'\i}as-P{\'e}rez}},
  \bibinfo{author}{G.~{Maggio}}, \bibinfo{author}{D.~{Maino}},
  \bibinfo{author}{N.~{Mandolesi}}, \bibinfo{author}{A.~{Mangilli}},
  \bibinfo{author}{A.~{Marchini}}, \bibinfo{author}{M.~{Maris}},
  \bibinfo{author}{P.~G. {Martin}}, \bibinfo{author}{M.~{Martinelli}},
  \bibinfo{author}{E.~{Mart{\'\i}nez-Gonz{\'a}lez}},
  \bibinfo{author}{S.~{Masi}}, \bibinfo{author}{S.~{Matarrese}},
  \bibinfo{author}{P.~{McGehee}}, \bibinfo{author}{P.~R. {Meinhold}},
  \bibinfo{author}{A.~{Melchiorri}}, \bibinfo{author}{J.~B. {Melin}},
  \bibinfo{author}{L.~{Mendes}}, \bibinfo{author}{A.~{Mennella}},
  \bibinfo{author}{M.~{Migliaccio}}, \bibinfo{author}{M.~{Millea}},
  \bibinfo{author}{S.~{Mitra}}, \bibinfo{author}{M.~A.
  {Miville-Desch{\^e}nes}}, \bibinfo{author}{A.~{Moneti}},
  \bibinfo{author}{L.~{Montier}}, \bibinfo{author}{G.~{Morgante}},
  \bibinfo{author}{D.~{Mortlock}}, \bibinfo{author}{A.~{Moss}},
  \bibinfo{author}{D.~{Munshi}}, \bibinfo{author}{J.~A. {Murphy}},
  \bibinfo{author}{P.~{Naselsky}}, \bibinfo{author}{F.~{Nati}},
  \bibinfo{author}{P.~{Natoli}}, \bibinfo{author}{C.~B. {Netterfield}},
  \bibinfo{author}{H.~U. {N{\o}rgaard-Nielsen}},
  \bibinfo{author}{F.~{Noviello}}, \bibinfo{author}{D.~{Novikov}},
  \bibinfo{author}{I.~{Novikov}}, \bibinfo{author}{C.~A. {Oxborrow}},
  \bibinfo{author}{F.~{Paci}}, \bibinfo{author}{L.~{Pagano}},
  \bibinfo{author}{F.~{Pajot}}, \bibinfo{author}{R.~{Paladini}},
  \bibinfo{author}{D.~{Paoletti}}, \bibinfo{author}{B.~{Partridge}},
  \bibinfo{author}{F.~{Pasian}}, \bibinfo{author}{G.~{Patanchon}},
  \bibinfo{author}{T.~J. {Pearson}}, \bibinfo{author}{O.~{Perdereau}},
  \bibinfo{author}{L.~{Perotto}}, \bibinfo{author}{F.~{Perrotta}},
  \bibinfo{author}{V.~{Pettorino}}, \bibinfo{author}{F.~{Piacentini}},
  \bibinfo{author}{M.~{Piat}}, \bibinfo{author}{E.~{Pierpaoli}},
  \bibinfo{author}{D.~{Pietrobon}}, \bibinfo{author}{S.~{Plaszczynski}},
  \bibinfo{author}{E.~{Pointecouteau}}, \bibinfo{author}{G.~{Polenta}},
  \bibinfo{author}{L.~{Popa}}, \bibinfo{author}{G.~W. {Pratt}},
  \bibinfo{author}{G.~{Pr{\'e}zeau}}, \bibinfo{author}{S.~{Prunet}},
  \bibinfo{author}{J.~L. {Puget}}, \bibinfo{author}{J.~P. {Rachen}},
  \bibinfo{author}{W.~T. {Reach}}, \bibinfo{author}{R.~{Rebolo}},
  \bibinfo{author}{M.~{Reinecke}}, \bibinfo{author}{M.~{Remazeilles}},
  \bibinfo{author}{C.~{Renault}}, \bibinfo{author}{A.~{Renzi}},
  \bibinfo{author}{I.~{Ristorcelli}}, \bibinfo{author}{G.~{Rocha}},
  \bibinfo{author}{C.~{Rosset}}, \bibinfo{author}{M.~{Rossetti}},
  \bibinfo{author}{G.~{Roudier}}, \bibinfo{author}{B.~{Rouill{\'e} d'Orfeuil}},
  \bibinfo{author}{M.~{Rowan-Robinson}}, \bibinfo{author}{J.~A.
  {Rubi{\~n}o-Mart{\'\i}n}}, \bibinfo{author}{B.~{Rusholme}},
  \bibinfo{author}{N.~{Said}}, \bibinfo{author}{V.~{Salvatelli}},
  \bibinfo{author}{L.~{Salvati}}, \bibinfo{author}{M.~{Sandri}},
  \bibinfo{author}{D.~{Santos}}, \bibinfo{author}{M.~{Savelainen}},
  \bibinfo{author}{G.~{Savini}}, \bibinfo{author}{D.~{Scott}},
  \bibinfo{author}{M.~D. {Seiffert}}, \bibinfo{author}{P.~{Serra}},
  \bibinfo{author}{E.~P.~S. {Shellard}}, \bibinfo{author}{L.~D. {Spencer}},
  \bibinfo{author}{M.~{Spinelli}}, \bibinfo{author}{V.~{Stolyarov}},
  \bibinfo{author}{R.~{Stompor}}, \bibinfo{author}{R.~{Sudiwala}},
  \bibinfo{author}{R.~{Sunyaev}}, \bibinfo{author}{D.~{Sutton}},
  \bibinfo{author}{A.~S. {Suur-Uski}}, \bibinfo{author}{J.~F. {Sygnet}},
  \bibinfo{author}{J.~A. {Tauber}}, \bibinfo{author}{L.~{Terenzi}},
  \bibinfo{author}{L.~{Toffolatti}}, \bibinfo{author}{M.~{Tomasi}},
  \bibinfo{author}{M.~{Tristram}}, \bibinfo{author}{T.~{Trombetti}},
  \bibinfo{author}{M.~{Tucci}}, \bibinfo{author}{J.~{Tuovinen}},
  \bibinfo{author}{M.~{T{\"u}rler}}, \bibinfo{author}{G.~{Umana}},
  \bibinfo{author}{L.~{Valenziano}}, \bibinfo{author}{J.~{Valiviita}},
  \bibinfo{author}{F.~{Van Tent}}, \bibinfo{author}{P.~{Vielva}},
  \bibinfo{author}{F.~{Villa}}, \bibinfo{author}{L.~A. {Wade}},
  \bibinfo{author}{B.~D. {Wandelt}}, \bibinfo{author}{I.~K. {Wehus}},
  \bibinfo{author}{M.~{White}}, \bibinfo{author}{S.~D.~M. {White}},
  \bibinfo{author}{A.~{Wilkinson}}, \bibinfo{author}{D.~{Yvon}},
  \bibinfo{author}{A.~{Zacchei}}, \bibinfo{author}{A.~{Zonca}},
\newblock \bibinfo{title}{{Planck 2015 results. XIII. Cosmological
  parameters}},
\newblock \bibinfo{journal}{\aap} \bibinfo{volume}{594} (\bibinfo{year}{2016})
  \bibinfo{pages}{A13}. \DOIprefix\doi{10.1051/0004-6361/201525830}.
  \href{http://arxiv.org/abs/1502.01589}{{\tt arXiv:1502.01589}}.
\bibitem[{{Fillmore} and {Goldreich}(1984)}]{fill84}
\bibinfo{author}{J.~A. {Fillmore}}, \bibinfo{author}{P.~{Goldreich}},
\newblock \bibinfo{title}{{Self-similar spherical voids in an expanding
  universe}},
\newblock \bibinfo{journal}{\apj} \bibinfo{volume}{281} (\bibinfo{year}{1984})
  \bibinfo{pages}{9--12}. \DOIprefix\doi{10.1086/162071}.
\bibitem[{{Cooray} and {Sheth}(2002)}]{coor02}
\bibinfo{author}{A.~{Cooray}}, \bibinfo{author}{R.~{Sheth}},
\newblock \bibinfo{title}{{Halo models of large scale structure}},
\newblock \bibinfo{journal}{\physrep} \bibinfo{volume}{372}
  (\bibinfo{year}{2002}) \bibinfo{pages}{1--129}.
  \DOIprefix\doi{10.1016/S0370-1573(02)00276-4}.
  \href{http://arxiv.org/abs/astro-ph/0206508}{{\tt arXiv:astro-ph/0206508}}.
\bibitem[{{Jose} et~al.(2013){Jose}, {Subramanian}, {Srianand }, and
  {Samui}}]{2013MNRAS.429.2333J}
\bibinfo{author}{C.~{Jose}}, \bibinfo{author}{K.~{Subramanian}},
  \bibinfo{author}{R.~{Srianand }}, \bibinfo{author}{S.~{Samui}},
\newblock \bibinfo{title}{{Spatial clustering of high-redshift Lyman-break
  galaxies}},
\newblock \bibinfo{journal}{\mnras} \bibinfo{volume}{429}
  (\bibinfo{year}{2013}) \bibinfo{pages}{2333--2350}.
  \DOIprefix\doi{10.1093/mnras/sts503}.
  \href{http://arxiv.org/abs/1208.2097}{{\tt arXiv:1208.2097}}.
\bibitem[{{Barkana} and {Loeb}(2001)}]{bark01}
\bibinfo{author}{R.~{Barkana}}, \bibinfo{author}{A.~{Loeb}},
\newblock \bibinfo{title}{{In the beginning: the first sources of light and the
  reionization of the universe}},
\newblock \bibinfo{journal}{\physrep} \bibinfo{volume}{349}
  (\bibinfo{year}{2001}) \bibinfo{pages}{125--238}.
  \DOIprefix\doi{10.1016/S0370-1573(01)00019-9}.
  \href{http://arxiv.org/abs/astro-ph/0010468}{{\tt arXiv:astro-ph/0010468}}.
\bibitem[{{Leitherer} et~al.(1999){Leitherer}, {Schaerer}, {Goldader},
  {Delgado}, {Robert}, {Kune}, {de Mello}, {Devost}, and {Heckman}}]{leith99}
\bibinfo{author}{C.~{Leitherer}}, \bibinfo{author}{D.~{Schaerer}},
  \bibinfo{author}{J.~D. {Goldader}}, \bibinfo{author}{R.~M.~G. {Delgado}},
  \bibinfo{author}{C.~{Robert}}, \bibinfo{author}{D.~F. {Kune}},
  \bibinfo{author}{D.~F. {de Mello}}, \bibinfo{author}{D.~{Devost}},
  \bibinfo{author}{T.~M. {Heckman}},
\newblock \bibinfo{title}{{Starburst99: Synthesis Models for Galaxies with
  Active Star Formation}},
\newblock \bibinfo{journal}{\apjs} \bibinfo{volume}{123} (\bibinfo{year}{1999})
  \bibinfo{pages}{3--40}. \DOIprefix\doi{10.1086/313233}.
  \href{http://arxiv.org/abs/astro-ph/9902334}{{\tt arXiv:astro-ph/9902334}}.
\bibitem[{{Thoul} and {Weinberg}(1996)}]{1996ApJ...465..608T}
\bibinfo{author}{A.~A. {Thoul}}, \bibinfo{author}{D.~H. {Weinberg}},
\newblock \bibinfo{title}{{Hydrodynamic Simulations of Galaxy Formation. II.
  Photoionization and the Formation of Low-Mass Galaxies}},
\newblock \bibinfo{journal}{\apj} \bibinfo{volume}{465} (\bibinfo{year}{1996})
  \bibinfo{pages}{608}. \DOIprefix\doi{10.1086/177446}.
  \href{http://arxiv.org/abs/astro-ph/9510154}{{\tt arXiv:astro-ph/9510154}}.
\bibitem[{{Samui} et~al.(2018){Samui}, {Srianand}, and
  {Subramanian}}]{samui18a}
\bibinfo{author}{S.~{Samui}}, \bibinfo{author}{R.~{Srianand}},
  \bibinfo{author}{K.~a. {Subramanian}},
\newblock \bibinfo{title}{{Probing feedback in high-$z$ galaxies using extended
  UV luminosity functions}},
\newblock \bibinfo{journal}{arXiv e-prints}  (\bibinfo{year}{2018})
  \bibinfo{pages}{arXiv:1805.05945}.
  \href{http://arxiv.org/abs/1805.05945}{{\tt arXiv:1805.05945}}.
\bibitem[{{Bromm} and {Loeb}(2002)}]{bromm02}
\bibinfo{author}{V.~{Bromm}}, \bibinfo{author}{A.~{Loeb}},
\newblock \bibinfo{title}{{The Expected Redshift Distribution of Gamma-Ray
  Bursts}},
\newblock \bibinfo{journal}{\apj} \bibinfo{volume}{575} (\bibinfo{year}{2002})
  \bibinfo{pages}{111--116}. \DOIprefix\doi{10.1086/341189}.
  \href{http://arxiv.org/abs/astro-ph/0201400}{{\tt arXiv:astro-ph/0201400}}.
\bibitem[{{Benson} et~al.(2002){Benson}, {Frenk}, and {Sharples}}]{benson02}
\bibinfo{author}{A.~J. {Benson}}, \bibinfo{author}{C.~S. {Frenk}},
  \bibinfo{author}{R.~M. {Sharples}},
\newblock \bibinfo{title}{{The Luminosity Functions and Stellar Masses of
  Galactic Disks and Spheroids}},
\newblock \bibinfo{journal}{\apj} \bibinfo{volume}{574} (\bibinfo{year}{2002})
  \bibinfo{pages}{104--113}. \DOIprefix\doi{10.1086/340925}.
  \href{http://arxiv.org/abs/astro-ph/0203464}{{\tt arXiv:astro-ph/0203464}}.
\bibitem[{{Dijkstra} et~al.(2004){Dijkstra}, {Haiman}, {Rees}, and
  {Weinberg}}]{dijk04}
\bibinfo{author}{M.~{Dijkstra}}, \bibinfo{author}{Z.~{Haiman}},
  \bibinfo{author}{M.~J. {Rees}}, \bibinfo{author}{D.~H. {Weinberg}},
\newblock \bibinfo{title}{{Photoionization Feedback in Low-Mass Galaxies at
  High Redshift}},
\newblock \bibinfo{journal}{\apj} \bibinfo{volume}{601} (\bibinfo{year}{2004})
  \bibinfo{pages}{666--675}. \DOIprefix\doi{10.1086/380603}.
  \href{http://arxiv.org/abs/astro-ph/0308042}{{\tt arXiv:astro-ph/0308042}}.
\bibitem[{{Kang} et~al.(2002){Kang}, {Jones}, and {Gieseler}}]{kangj02}
\bibinfo{author}{H.~{Kang}}, \bibinfo{author}{T.~W. {Jones}},
  \bibinfo{author}{U.~D.~J. {Gieseler}},
\newblock \bibinfo{title}{{Numerical Studies of Cosmic-Ray Injection and
  Acceleration}},
\newblock \bibinfo{journal}{\apj} \bibinfo{volume}{579} (\bibinfo{year}{2002})
  \bibinfo{pages}{337--358}. \DOIprefix\doi{10.1086/342724}.
  \href{http://arxiv.org/abs/astro-ph/0207410}{{\tt arXiv:astro-ph/0207410}}.
\bibitem[{{Kang} and {Jones}(2003)}]{kangj03}
\bibinfo{author}{H.~{Kang}}, \bibinfo{author}{T.~W. {Jones}},
\newblock \bibinfo{title}{{Cosmic Ray Acceleration at Quasi-Parallel Plane
  Shocks}},
\newblock \bibinfo{journal}{International Cosmic Ray Conference}
  \bibinfo{volume}{4} (\bibinfo{year}{2003}) \bibinfo{pages}{2039}.
\bibitem[{{Kang} and {Jones}(2005)}]{kangj05}
\bibinfo{author}{H.~{Kang}}, \bibinfo{author}{T.~W. {Jones}},
\newblock \bibinfo{title}{{Efficiency of Nonlinear Particle Acceleration at
  Cosmic Structure Shocks}},
\newblock \bibinfo{journal}{\apj} \bibinfo{volume}{620} (\bibinfo{year}{2005})
  \bibinfo{pages}{44--58}. \DOIprefix\doi{10.1086/426855}.
  \href{http://arxiv.org/abs/astro-ph/0410724}{{\tt arXiv:astro-ph/0410724}}.
\bibitem[{{Furlanetto} and {Loeb}(2001)}]{furl01}
\bibinfo{author}{S.~R. {Furlanetto}}, \bibinfo{author}{A.~{Loeb}},
\newblock \bibinfo{title}{{Intergalactic Magnetic Fields from Quasar
  Outflows}},
\newblock \bibinfo{journal}{\apj} \bibinfo{volume}{556} (\bibinfo{year}{2001})
  \bibinfo{pages}{619--634}. \DOIprefix\doi{10.1086/321630}.
  \href{http://arxiv.org/abs/astro-ph/0102076}{{\tt arXiv:astro-ph/0102076}}.
\bibitem[{{Navarro} et~al.(1997){Navarro}, {Frenk}, and {White}}]{navar97}
\bibinfo{author}{J.~F. {Navarro}}, \bibinfo{author}{C.~S. {Frenk}},
  \bibinfo{author}{S.~D.~M. {White}},
\newblock \bibinfo{title}{{A Universal Density Profile from Hierarchical
  Clustering}},
\newblock \bibinfo{journal}{\apj} \bibinfo{volume}{490} (\bibinfo{year}{1997})
  \bibinfo{pages}{493--508}. \DOIprefix\doi{10.1086/304888}.
  \href{http://arxiv.org/abs/astro-ph/9611107}{{\tt arXiv:astro-ph/9611107}}.
\bibitem[{{Mori} et~al.(2002){Mori}, {Ferrara}, and {Madau}}]{mori02}
\bibinfo{author}{M.~{Mori}}, \bibinfo{author}{A.~{Ferrara}},
  \bibinfo{author}{P.~{Madau}},
\newblock \bibinfo{title}{{Early Metal Enrichment by Pregalactic Outflows. II.
  Three-dimensional Simulations of Blow-Away}},
\newblock \bibinfo{journal}{\apj} \bibinfo{volume}{571} (\bibinfo{year}{2002})
  \bibinfo{pages}{40--55}. \DOIprefix\doi{10.1086/339913}.
  \href{http://arxiv.org/abs/astro-ph/0106107}{{\tt arXiv:astro-ph/0106107}}.
\bibitem[{{Uhlig} et~al.(2012){Uhlig}, {Pfrommer}, {Sharma}, {Nath},
  {En{\ss}lin}, and {Springel}}]{2012MNRAS.423.2374U}
\bibinfo{author}{M.~{Uhlig}}, \bibinfo{author}{C.~{Pfrommer}},
  \bibinfo{author}{M.~{Sharma}}, \bibinfo{author}{B.~B. {Nath}},
  \bibinfo{author}{T.~A. {En{\ss}lin}}, \bibinfo{author}{V.~{Springel}},
\newblock \bibinfo{title}{{Galactic winds driven by cosmic ray streaming}},
\newblock \bibinfo{journal}{\mnras} \bibinfo{volume}{423}
  (\bibinfo{year}{2012}) \bibinfo{pages}{2374--2396}.
  \DOIprefix\doi{10.1111/j.1365-2966.2012.21045.x}.
  \href{http://arxiv.org/abs/1203.1038}{{\tt arXiv:1203.1038}}.
\bibitem[{{Fujita} and {Mac Low}(2018)}]{2018MNRAS.477..531F}
\bibinfo{author}{A.~{Fujita}}, \bibinfo{author}{M.-M. {Mac Low}},
\newblock \bibinfo{title}{{Cosmic ray driven outflows in an ultraluminous
  galaxy}},
\newblock \bibinfo{journal}{\mnras} \bibinfo{volume}{477}
  (\bibinfo{year}{2018}) \bibinfo{pages}{531--538}.
  \DOIprefix\doi{10.1093/mnras/sty715}.
  \href{http://arxiv.org/abs/1804.10302}{{\tt arXiv:1804.10302}}.
\bibitem[{{Beck}(2015)}]{rbeck15}
\bibinfo{author}{R.~{Beck}},
\newblock \bibinfo{title}{{Magnetic fields in spiral galaxies}},
\newblock \bibinfo{journal}{\aapr} \bibinfo{volume}{24} (\bibinfo{year}{2015})
  \bibinfo{pages}{4}. \DOIprefix\doi{10.1007/s00159-015-0084-4}.
  \href{http://arxiv.org/abs/1509.04522}{{\tt arXiv:1509.04522}}.
\bibitem[{{Montero-Dorta} and {Prada}(2009)}]{monter09}
\bibinfo{author}{A.~D. {Montero-Dorta}}, \bibinfo{author}{F.~{Prada}},
\newblock \bibinfo{title}{{The SDSS DR6 luminosity functions of galaxies}},
\newblock \bibinfo{journal}{\mnras} \bibinfo{volume}{399}
  (\bibinfo{year}{2009}) \bibinfo{pages}{1106--1118}.
  \DOIprefix\doi{10.1111/j.1365-2966.2009.15197.x}.
  \href{http://arxiv.org/abs/0806.4930}{{\tt arXiv:0806.4930}}.
\bibitem[{{Sasaki}(1994)}]{sasaki94}
\bibinfo{author}{S.~{Sasaki}},
\newblock \bibinfo{title}{{Formation rate of bound objects in the hierarchical
  clustering model}},
\newblock \bibinfo{journal}{\pasj} \bibinfo{volume}{46} (\bibinfo{year}{1994})
  \bibinfo{pages}{427--430}.
\bibitem[{{Samui} et~al.(2009){Samui}, {Srianand}, and {Subramanian}}]{samui09}
\bibinfo{author}{S.~{Samui}}, \bibinfo{author}{R.~{Srianand}},
  \bibinfo{author}{K.~{Subramanian}},
\newblock \bibinfo{title}{{Understanding the redshift evolution of the
  luminosity functions of Lyman {$\alpha$} emitters}},
\newblock \bibinfo{journal}{\mnras} \bibinfo{volume}{398}
  (\bibinfo{year}{2009}) \bibinfo{pages}{2061--2068}.
  \DOIprefix\doi{10.1111/j.1365-2966.2009.15245.x}.
  \href{http://arxiv.org/abs/0906.2312}{{\tt arXiv:0906.2312}}.
\bibitem[{{Alpaslan} et~al.(2015){Alpaslan}, {Driver}, {Robotham},
  {Obreschkow}, {Andrae}, {Cluver}, {Kelvin}, {Lange}, {Owers}, {Taylor},
  {Andrews}, {Bamford}, {Bland-Hawthorn}, {Brough}, {Brown}, {Colless},
  {Davies}, {Eardley}, {Grootes}, {Hopkins}, {Kennedy}, {Liske},
  {Lara-L{\'o}pez}, {L{\'o}pez-S{\'a}nchez}, {Loveday}, {Madore}, {Mahajan},
  {Meyer}, {Moffett}, {Norberg}, {Penny}, {Pimbblet}, {Popescu}, {Seibert}, and
  {Tuffs}}]{2015MNRAS.451.3249A}
\bibinfo{author}{M.~{Alpaslan}}, \bibinfo{author}{S.~{Driver}},
  \bibinfo{author}{A.~S.~G. {Robotham}}, \bibinfo{author}{D.~{Obreschkow}},
  \bibinfo{author}{E.~{Andrae}}, \bibinfo{author}{M.~{Cluver}},
  \bibinfo{author}{L.~S. {Kelvin}}, \bibinfo{author}{R.~{Lange}},
  \bibinfo{author}{M.~{Owers}}, \bibinfo{author}{E.~N. {Taylor}},
  \bibinfo{author}{S.~K. {Andrews}}, \bibinfo{author}{S.~{Bamford}},
  \bibinfo{author}{J.~{Bland-Hawthorn}}, \bibinfo{author}{S.~{Brough}},
  \bibinfo{author}{M.~J.~I. {Brown}}, \bibinfo{author}{M.~{Colless}},
  \bibinfo{author}{L.~J.~M. {Davies}}, \bibinfo{author}{E.~{Eardley}},
  \bibinfo{author}{M.~W. {Grootes}}, \bibinfo{author}{A.~M. {Hopkins}},
  \bibinfo{author}{R.~{Kennedy}}, \bibinfo{author}{J.~{Liske}},
  \bibinfo{author}{M.~A. {Lara-L{\'o}pez}}, \bibinfo{author}{{\'A}.~R.
  {L{\'o}pez-S{\'a}nchez}}, \bibinfo{author}{J.~{Loveday}},
  \bibinfo{author}{B.~F. {Madore}}, \bibinfo{author}{S.~{Mahajan}},
  \bibinfo{author}{M.~{Meyer}}, \bibinfo{author}{A.~{Moffett}},
  \bibinfo{author}{P.~{Norberg}}, \bibinfo{author}{S.~{Penny}},
  \bibinfo{author}{K.~A. {Pimbblet}}, \bibinfo{author}{C.~C. {Popescu}},
  \bibinfo{author}{M.~{Seibert}}, \bibinfo{author}{R.~{Tuffs}},
\newblock \bibinfo{title}{{Galaxy And Mass Assembly (GAMA): trends in galaxy
  colours, morphology, and stellar populations with large-scale structure,
  group, and pair environments}},
\newblock \bibinfo{journal}{\mnras} \bibinfo{volume}{451}
  (\bibinfo{year}{2015}) \bibinfo{pages}{3249--3268}.
  \DOIprefix\doi{10.1093/mnras/stv1176}.
  \href{http://arxiv.org/abs/1505.05518}{{\tt arXiv:1505.05518}}.

\end{thebibliography}
\bibliographystyle{elsarticle-num-names}

\end{document}